\documentclass[a4paper,11pt]{article}

\usepackage{jcappub} 
\usepackage{graphicx}
\usepackage{amssymb}
\usepackage{filecontents}
\usepackage{amsmath}
\usepackage{color}
\usepackage{float}
\usepackage{textcomp}
\usepackage{bm}
\usepackage{natbib}

\usepackage{hyperref}
\usepackage{esint}	
\usepackage{mathtools}

\usepackage{xcolor}

\author[a,1]{Elham Nazari,\note{Corresponding author.}}
\author[b]{Samik Mitra,}
\author[c]{Shahram Abbassi,}
\author[b]{Santabrata Das}

\affiliation[a]{Department of Physics, Faculty of Science, Ferdowsi University of Mashhad, P.O. Box 1436, Mashhad, Iran}
\affiliation[b]{Department of Physics, Indian Institute of Technology Guwahati, Assam, India, 781039}
\affiliation[c]{Department of Physics \& Astronomy, University of Western Ontario, London, ON, N6A 3K7, Canada}

\emailAdd{elham.nazari@mail.um.ac.ir}
\emailAdd{m.samik@iitg.ac.in}
\emailAdd{sabbassi@uwo.ca}
\emailAdd{sbdas@iitg.ac.in}

\title{Accretion flows around spinning compact objects in the post-Newtonian regime}

\abstract{
We present the structure of a low angular momentum accretion flows around rotating compact objects incorporating relativistic corrections up to the leading post-Newtonian order. To begin with, we formulate the governing post-Newtonian hydrodynamic equations for the mass and energy-momentum flux without imposing any symmetries. However, for the sake of simplicity, we consider the flow to be stationary, axisymmetric, and inviscid. Toward this, we adapt the polytropic equation of state (EoS) and analyze the vertically integrated accretion flow confined to the equatorial plane.  It is shown that the spin-orbit effects manifest themselves in the accretion dynamics. In the present analysis, we focus on global transonic accretion solutions, where a subsonic flow enters far away from the compact object and gradually gains radial velocity as it moves inwards. Thus, the flow becomes supersonic after reaching a certain radius, known as the critical point. To better understand the transonic solutions and examine the effect of post-Newtonian corrections, we classify the post-Newtonian equations into semi-relativistic (SR), semi-Newtonian (SN), and non-relativistic (NR) limits and compare the accretion solutions and their corresponding flow variables. With these, we find that SR and SN flow are in good agreement all throughout, although they deviate largely from the NR ones. Interestingly, the density profile seems to follow the profile $\rho \propto r^{-3/2}$ in the post-Newtonian regime. The present study has the potential to connect Newtonian and GR descriptions of accretion dynamics. } 
\keywords{Accretion -- post-Newtonian -- gravity -- hydrodynamics}

\begin{document}

\maketitle
\flushbottom

\section{Introduction}

Accretion flows around compact objects, such as black holes (BHs) and neutron stars, have garnered significant attention in astrophysics due to their pivotal role in a wide range of high-energy astrophysical phenomena \textit{i.e.,} emergent electromagnetic radiations from quasars, active galactic nuclei, and BH X-ray binaries. The underlying description through which one studies the accretion process is the hydrodynamic or magneto-hydrodynamic flow of matter in the surroundings of a central compact object.  In this regard, Shakura \& Sunyaev \cite[hereby known as SS73 model]{1973A&A....24..337S} introduced the concept of a geometrically thin and optically thick accretion disk, providing a foundation for understanding the observed properties of accreting systems. This thin disk model is particularly applicable in regions of the accretion flows located farther from the central black hole, where the accretion rate is relatively high and the matter has a Keplerian angular momentum distribution. \cite{1973blho.conf..343N} extended the SS73 model to incorporate relativistic effects near the black hole, such as the strong gravitational field and the associated frame-dragging effects caused by the black hole's rotation. Novikov and Thorne's \cite[]{1973blho.conf..343N} relativistic solution provided valuable insights for subsequent relativistic studies considering the effects of black hole spin, disk thickness, and radiative processes \cite[]{1974ApJ...191..499P,1975ApJ...202..788C}. 

However, to describe the innermost regions of an accretion disk, alternative models have been proposed. The advection-dominated accretion flow (ADAF) \cite{1994ApJ...428L..13N} is one such model that has gained acceptance. ADAF emphasizes the inefficient cooling due to advection, where a significant fraction of the energy generated by inflowing matter remains trapped within the flow rather than being radiated away. Such an accretion flow model has been widely applied to the low luminous AGNs (\textit{i.e.,} Sgr A*) and BH binaries in their hard and quiescent states \cite[and references therein]{2008NewAR..51..733N,2014ARA&A..52..529Y}. Motivated by this, we study the low angular momentum advective accretion flows around compact objects in this work. For such choices of angular momentum distribution, the circularization radius \cite{2002apa..book.....F} comes out to be of the order of a few hundred $r_{\rm g}$, the gravitational radius of the body. Over the years, semi-analytic studies of low angular momentum transonic flows, e.g., for hydro case \cite{2007MNRAS.376.1659D,2022MNRAS.514.1940D}, for MHD \cite{2022MNRAS.516.5092M} reveal different accretion regimes characterized by smooth Bondi-like flows, standing accretion shocks or the formation of circularized tori \cite{2023A&A...678A.141O}.

In the subsequent analysis, we focus on the gravity model, which is best to address the dynamical structure of an accretion flow. In reality, when the inflowing matter approaches the horizon (or surface for supermassive stars or neutron stars), the general relativistic (GR) effects become dominant, and due to a high degree of non-linearity, it is generally difficult to solve the problem trivially. To avoid such complexity, most of the studies of the accretion flow around the compact object were confined to the Newtonian regime, where the gravitational effect is taken into account using effective potentials. However, the Newtonian potential was not sufficient enough to explain the gravity, particularly around black holes, and the pseudo-Newtonian potentials (\cite{1980A&A....88...23P}; hereby PW80) gained much appreciation due to its simplicity. For a non-rotating BH, PW80 successfully describes the GR features. In a recent work, \cite{2018PhRvD..98h3004D} pointed out several limitations of such a pseudo-Newtonian framework, particularly for the rotating BH case \cite{1992MNRAS.256..300C,1996ApJ...461..565A,2002ApJ...581..427M,2006MNRAS.369..976C,2007ApJ...667..367G,2014MNRAS.445.4463G}, and successfully obtained the first global transonic accretion solutions around a rotating BH for a wide range of spin\footnote{The term ``spin'' is utilized to describe the intrinsic angular momentum of a rotating body.} variations ($-1\le a_{\rm k} \le 1$, where $a_{\rm k}$ is the spin of the BH) following a complete GR treatment.

An approach called numerical relativity has been established to study realistic systems in GR.
However, to examine complicated realistic systems like accretion flows, sophisticated numerical simulations are often needed (see for instance \cite{ 2003ApJ...589..444G,2016ApJS..225...22W,2017ComAC...4....1P} and references therein), which are intractable in some studies  \cite{2023PhRvX..13b1035S}. From this perspective, analytical methods are still of particular importance. With their help, one can investigate the rich astrophysics of accretion disks \cite{2023PhRvX..13b1035S}.

Another approach to study realistic systems in GR analytically is the approximate method. One of the most successful approximations to GR based on the slow-motion condition\footnote{It means that the motions in physical systems are slow compared to the speed of light, $c$.} and the weak-field limit is the post-Newtonian theory. It has been proven that this approximation, in a systematic way, is effective in describing gravitational physics \cite{poisson2014gravity,1993tegp.book.....W,2011PNAS..108.5938W}. Unlike exact GR solutions, post-Newtonian results are not restricted to a specific symmetry. By using this approximation and systematically expanding the results to higher post-Newtonian order, one can describe an asymmetric and complex relativistic system to the desired accuracy. In the legendary EIH paper \cite{1938AnMat..39...65E}, Einstein, Infeld, and Hoffmann first introduced the $N$-body equations of motion in the weak field and slow-motion limits. The post-Newtonian hydrodynamics was then developed by \cite{1964tstg.book.....F,1965ApJ...142.1488C,1967ApJ...148..621C,1969ApJ...158...45C,1969ApJ...158...55C,1970ApJ...160..153C}. Applying this approximation, including relativistic effects, GR is experimentally tested, see, \textit{e.g.}, \cite{1971ApJ...163..595T,1971ApJ...163..611W,1971ApJ...169..125W,1987thyg.book...80W,1994reco.conf...83W}. This method is widely used to study binary systems \cite{1989AIHPA..50..377B,1991ApJ...366..501D,1995PhRvD..51.5360B,2006LRR.....9....4B} and has a significant role in the discovery of gravitational waves, cf. \textit{e.g.}, \cite{2017PhRvL.119p1101A}. Furthermore, it is utilized in various contexts, such as dynamical instability of neutron-star binaries \cite{1996PhRvD..54.3958L,2000PhRvD..62f4012F}, gravitational instabilities in hydrodynamic fluid \cite{2017ApJ...839...75N,2018ApJ...865...71K}, post-Newtonian effects in magnetohydrodynamics \cite{1971ApJ...164..589G,2018ApJ...868...98N,2020ApJ...899...59H} and its role in the gravitational radiation by magnetic field of magnetars \cite{2020MNRAS.498..110N}. Post-Newtonian effects on accretion flow are also studied in several perspectives \cite{1997A&A...324..829D,1997MNRAS.285..394I,2015PhRvD..91b4039J,2016PhRvD..94l4041K,2021PhRvD.104b4056K}.

Following the above-mentioned details, we derive the post-Newtonian hydrodynamic equations governing the accretion dynamics around a rotating compact body. Our derivation is based on the modern post-Newtonian approach introduced by \cite{poisson2014gravity}. This formalism is developed by \cite{1976ApJ...210..764W}, and extended by \cite{1996PhRvD..54.4813W} and  \cite{2000PhRvD..62l4015P,2002PhRvD..65j4008P}. This approach is closely related to the ``post-Minkowskian'' framework of Blanchet, Damour, and Iyer. We refer the interested reader to \cite{2014LRR....17....2B}. 
The main goal is to study accretion flows around spinning \ compact objects to the first post-Newtonian (1\tiny PN \normalsize) order.
This study can be treated as a bridge between Newtonian and GR descriptions of accretion flows, especially in important situations involving spin effects\footnote{ Recently, regarding the observational evidence that black holes can be rotating, it has been widely accepted that astrophysical black holes have non-zero spin, \textit{e.g.}, see  \cite{2009ApJ...697..900M,2010MNRAS.403L..74K,2011CQGra..28k4009M,2013Sci...339...49M}.}. A relativistic phenomenon is obtained here\textemdash the spin-orbit interaction. This effect is not present in Newtonian gravity.
The 1\tiny PN \normalsize effects of spin have been studied by several authors from different perspectives, see \cite{1995PhRvD..52..821K,2005PhRvD..71h4027W} and reference therein. It is shown that spin-orbit accelerations as well as spin-spin accelerations play a role in the equations of motion of the $N$-body system in post-Newtonian gravity \cite{poisson2014gravity,1993tegp.book.....W}. 
In this work, the footprint of the spin-spin effect is also observed in the higher post-Newtonian order. To fully analyze this effect, one needs to completely examine the second post-Newtonian corrections which is beyond the scope of the present work.
Also, as expected, this approximation reproduces the predictions given in Newtonian gravity in the leading order.

It should be noted that the convergence properties of the post-Newtonian approximation are not clear well; and due to this lack of knowledge, it is not known a priori to which region of a relativistic system this approximation is applicable and where it is no longer valid \cite{2011PNAS..108.5938W}. Another point to be investigated here is the approximate representation of this region in accretion flows around spinning compact objects.  To do this, we compare our findings with those of GR. Despite the limitations of the post-Newtonian approximation, namely weak-field limit and slow-motion condition, the results are reliable even where these constraints are not met. In fact, the post-Newtonian prescription can be effective even beyond the realm of its validity, \textit{i.e.,} $r<10\,r_g$. Here, $r_g \coloneqq \frac{G\, M}{c^2}$ is the gravitational radius of a body of mass $M$.
This ``astounding'' property of post-Newtonian gravity is also observed in the context of gravitational wave studies \cite{2011PNAS..108.5938W}.  With all the above considerations, we solve the mass and energy-momentum conservation equations and obtain a complete set of global transonic accretion solutions around spinning compact objects in the post-Newtonian background. In doing so, we choose the ideal equation of state (IEoS) to describe thermodynamics. 

The paper is organized as follows. The post-Newtonian spacetime around a spinning body is introduced in Sec. \ref{sec2}. In Sec. \ref{sec3}, we describe in detail the method of deriving the hydrodynamic equations governing a fluid embedded in this spacetime.  Sec. \ref{sec4} shows a detailed discussion on the governing flow equations in the post-Newtonian spacetime of a spinning compact body. The final results are rendered in the reference frame comoving with the fluid element. 
We attempt to evaluate the transonic properties of the accretion flows in post-Newtonian gravity in Sec. \ref{sec5}. Our conclusions are presented in Sec. \ref{sec6}, while Appendices \ref{app1}-\ref{app3} summarize the necessary relations needed during computations, and Appendix \ref{discussion} discusses the distribution of angular momentum and the disk luminosity. Appendix \ref{viscous} also provides a discussion of angular momentum transport in viscous accretion disks.

A word about conventions: In the post-Newtonian framework, each order $c^{-2}$ is considered a post-Newtonian correction. To derive the relations, we expand the result in powers of $c$ and keep only the terms up to the required orders. Moreover, the viscous transport of angular momentum is not taken into account in this paper. Here, Latin and Greek indices run over the
values $\lbrace1, 2, 3\rbrace$ and $\lbrace 0, 1, 2, 3 \rbrace$, respectively.

\section{Post-Newtonian spacetime}\label{sec2}

We first need to be armed with an appropriate metric that correctly describes the spacetime around a spinning body in the post-Newtonian regime. In this work, we consider an accretion disk with negligible self-gravity. So, the curvature of the spacetime is only affected by the central spinning body. In the post-Newtonian framework, doing very detailed calculations, this metric is obtained; and here, we apply the results given in \cite{poisson2014gravity}. We do not repeat this tedious derivation and refer the interested reader to this standard textbook.

It is shown that the metric of the spacetime outside a single spinning body up to the 1\tiny PN \normalsize order is given by,
\begin{subequations}
\begin{align}
\label{g_00}
&g_{00}=-1+\frac{2}{c^2}\frac{G\,M}{r}-\frac{2}{c^4}\Big(\frac{G\,M}{r}\Big)^2+O(c^{-6}),\\
&g_{0j}=\frac{2}{c^3}\frac{G(\bm{x}\times\bm{S})^j}{r^3}+O(c^{-5}),\\
\label{g_jk}
&g_{jk}=\Big(1+\frac{2}{c^2}\frac{G\,M}{r}\Big)\delta_{jk}+O(c^{-4}),
\end{align}
\end{subequations}
where $M$ and $\bm{S}$ are the mass and spin vector of the body, respectively. Also, $\bm{x}$ is the position vector of the field point outside the body. Its component notation is $x^j\coloneqq \big(x, y, z\big)$. Here, $r \coloneqq |\bm{x}|$ and $\delta_{jk}$ is the Kronecker delta. 
In this spacetime, the metric determinant is given by,
\begin{align}\label{det_g}
-g=1+\frac{4}{c^2}\frac{G\,M}{r}+O(c^{-4}).
\end{align} 
We examine the behavior of relativistic and Newtonian fluids in this spacetime.

As seen, in Eqs. \eqref{g_00}-\eqref{g_jk}, the post-Newtonian expansion of the time-time, time-space, and space-space components is truncated to $O(c^{-4})$, $O(c^{-3})$, and $O(c^{-2})$, respectively. In fact, having these orders for the metric components, one can approximately investigate the general relativistic aspects of a system to the 1\tiny PN \normalsize order \cite{poisson2014gravity}.    
It should be noticed that this metric describes the spacetime outside the spinning body approximately, and it is suitable for studying the 1\tiny PN \normalsize corrections of the curved spacetime to the studied system. Of course, it is also assumed that the fluid system around the spinning body does not generate the metric. Then, we do not expect this metric to provide us with results obtained from the exact Kerr background. 
In other words, this metric can be used to study the behavior of a system close to the central body, but not too close where the approximation breaks down and full general relativity with the exact Kerr solution is needed, somewhere between the spacetime associated with the Newtonian field, far from the gravitational body, and Kerr spacetime very close to it.

Now, an important question may be raised as to what radius, this approximation works and, in principle, in what region around the central object, it can be applied. 
In fact, in the accretion disk system, there is a radius where the velocity of the fluid element, as well as the gravitational fields, can be large so that the post-Newtonian gravity is no longer valid. On the other hand, due to our lack of knowledge of the convergence properties of this approximation, this radius is not known a priori \cite{2011PNAS..108.5938W}. However, to answer this question and obtain the post-Newtonian validity domain, we apply the following scheme.

We analyze the aforementioned post-Newtonian metric and compare the post-Newtonian terms of its components with the Newtonian one. We know that the second term in Eq. \eqref{g_00} is a Newtonian term. Also, as mentioned earlier, the third term in this relation is the first post-Newtonian correction. It is then obvious that the ratio of the third term to the second is too small (the weak-field limit $\frac{G\,M}{r\,c^2}\ll 1$). Therefore, from this comparison, the constraint $r\gg r_g$ is deduced. In a similar fashion, comparing the post-Newtonian term in the time-space component of the metric with the Newtonian term in the time-time component also reveals that $r\gg\frac{S}{c\,M}$. This limit indeed demands a weaker constraint on $r$. To see this fact, we consider an extreme Kerr black hole with $S=\frac{G\, M^2}{c}$. Inserting this value into the condition $r\gg\frac{S}{c\,M}$, we arrive at $r\gg r_g$ obtained before. So, the condition resulting from the comparison of the post-Newtonian and Newtonian terms of the time-time component places a tighter bound on $r$ and automatically fulfills the condition $r\gg\frac{S}{c\,M}$. Based on this discussion, the post-Newtonian metric introduced here can describe the spacetime at the radius where $r\gg r_g$. In this work, we consider that this radius is an order of magnitude larger than $r_g$ so that this condition is satisfied. Henceforth, it is called the post-Newtonian radius, whose value is $r_{\text{PN}}=10\,r_g$.  Strictly speaking, the following analyses, restricted to the 1\tiny PN \normalsize order, are reliable from very far radii, $r_{\infty}$, up to $r_{\text{PN}}$. 
In the following, comparing the effective potential obtained from the exact Kerr spacetime with the post-Newtonian potential, this statement is objectively supported, cf. Fig. \ref{fig:pot} which demonstrates this fact well.
To investigate a system at inner radii, \textit{i.e.,} $r<r_{\text{PN}}$, one needs to take into account the exact relativistic corrections and apply the Kerr metric. This is indeed one of the main shortcomings of the post-Newtonian approximation. Furthermore, in the post-Newtonian framework, to improve the measurements in the post-Newtonian zone, \textit{i.e.,} $r_{\text{PN}} \leqslant r \leqslant r_{\infty}$, one should consider higher corrections and utilize the post-Newtonian expansion of the metric containing at least the 2\tiny PN \normalsize corrections. As a first step toward the study of accretion flows around spinning bodies in post-Newtonian gravity, we restrict ourselves to the 1\tiny PN \normalsize order and apply Eqs. \eqref{g_00}-\eqref{g_jk} in the current paper.

\section{Governing hydrodynamic equations up to the 1\tiny PN \normalsize order}\label{sec3}

In this section, we study the hydrodynamic equations governing the behavior of a perfect fluid in post-Newtonian spacetime. To prepare the way for the subsequent calculations, we introduce the contravariant components of the metric \eqref{g_00}-\eqref{g_jk} as well as the corresponding Christoffel symbols in Appx. \ref{app1}. 

\subsection{Relativistic perfect fluid}

For the matter part of the current study, as the simplest case, it is assumed that the fluid system is neutral and perfect. 
The energy-momentum tensor of the perfect fluid is given by
\begin{align}\label{T_fluid}
T^{\alpha\beta}=\left(\rho+\frac{\epsilon}{c^2}+\frac{p}{c^2}\right)u^{\alpha}u^{\beta}+pg^{\alpha\beta},
\end{align}
where $\rho$ is the proper rest-mass density, $\epsilon$ is the proper internal energy density, $p$ is thermal pressure, and $u^{\alpha}=\gamma(c,v^j)$ is the four-velocity field. Here, $\gamma=u^0/c$. In this spirit, we then neglect viscous effects, heat fluxes, and magnetic and radiation fields in the accretion disks throughout this study (for further details on viscosity in accretion disks, see Appx. \ref{viscous}).

As mentioned, the post-Newtonian approximation relies on slow-motion and weak-field limits. In this framework, the nature of gravity is investigated in the weak-field situation, \textit{i.e.,} 
\begin{align}\label{condi2}
\frac{G\,M}{c^2 r}\ll 1,
\end{align}
and it is assumed that the matter distribution moves slowly, namely slow-motion condition as
\begin{align}\label{condi1}
\frac{v^2}{c^2}\ll 1.
\end{align}
According to the condition \eqref{condi2}, the hydrostatic equilibrium, and the relation between pressure and energy, we also have  
\begin{align}
\frac{p}{\rho^*c^2}\sim\frac{\Pi}{ c^2}\ll 1.
\end{align}
Here, $\Pi=\epsilon/\rho$ is the internal energy of the fluid element divided by its mass.
We assign the label $O(c^{-2})$ to denote the order of the smallness of the above quantities. 
Here, by relativistic fluid, we mean a system for which the above conditions 
are established\footnote{This fluid is only subject to the gravitational field of the central black hole. Since this fluid system is assumed to have negligible self-gravity, we do not need to make any conditions on its gravitational field. In fact, it is automatically removed from the set of equations.} and the linear combination of these terms plays a role in the hydrodynamic equations. In fact, these ratios being the 1\tiny PN \normalsize corrections, $O(c^{-2})$, are kept as the relativistic effects in the equation of motions. To study a system that is more relativistic, in addition to leading-order pieces, \textit{i.e.,} Newtonian terms, and the 1\tiny PN \normalsize corrections, higher powers of these terms and their combinations, \textit{i.e.,} $O(c^{-n})$, where $n>2$ must be taken into account. In line with our goal in this paper, we drop these higher-order terms and consider the role of $O(c^{-2})$ terms.

\subsection{Relativistic Hydrodynamics}

As we know, two conservation laws mainly represent the general-relativistic hydrodynamic equations. The first one is the conservation of the rest mass expressing that
\begin{align}
\nabla_{\mu}\left(\rho u^{\mu}\right)=0, 
\end{align}
and the second is the conservation of the energy-momentum tensor given by,
\begin{align}
\nabla_{\beta}T^{\alpha\beta}=0. 
\end{align}
It can easily be shown that these conservation relations are simplified as  
\begin{align}
\label{cont_eq}
\nabla_{\mu}\left(\rho u^{\mu}\right)=\partial_t \rho^*+\partial_j\big(\rho^* v^j\big)=0,
\end{align}
and
\begin{align}\label{con_EM}
\nabla_{\beta}T^{\alpha\beta}=\partial_{\beta}\big(\sqrt{-g}T^{\alpha\beta}\big)+\Gamma^{\alpha}_{\beta \mu}\big(\sqrt{-g}T^{\beta\mu}\big)=0,
\end{align}
respectively. To obtain the latter equation, we use the relation $\Gamma^\mu_{\mu\beta}=\big(-g\big)^{-1/2}\partial_\beta\big(-g\big)^{1/2}$.  
Here, $\rho^*$ is the rescaled mass density defined as 
\begin{align}\label{rho*}
\rho^*=\sqrt{-g}\gamma\rho.
\end{align}
Hereafter, unless otherwise specified, we use $\rho^*$ in the context of the post-Newtonian gravity. Eqs. \eqref{cont_eq} and \eqref{con_EM} are in fact the equations governing the behavior of a generic fluid in a general curved spacetime. To close the system of equations, one also needs to determine the EoS of the fluid, known as a closure relation in the set of hydrodynamics equations.

In the following computations, we need to know the quantity $\gamma$ up to $O(c^{-2})$. To do so, we use the normalization condition $g_{\alpha\beta}u^{\alpha}u^{\beta}=-c^2$. After inserting the definition of the four-velocity field and the metric components described by \eqref{g_00}-\eqref{g_jk} and applying the conditions \eqref{condi1}, with some manipulations, one can arrive at
\begin{align}\label{gamma}
\gamma=1+\frac{1}{c^2}\frac{G\,M}{r}+\frac{1}{2c^2}v^2+O(c^{-4}).
\end{align}
With Eqs. \eqref{det_g} and \eqref{gamma} in hand, we return to the definition \eqref{rho*} and find
\begin{align}\label{rho*PN}
\rho^*=\rho\Big(1+\frac{1}{c^2}\frac{3\,G\,M}{r}+\frac{1}{2c^2}v^2\Big)+O(c^{-4}),
\end{align}
after expanding the result in powers of $c$ and truncating it to $O(c^{-2})$.

We first turn to study the time component of the energy-momentum conservation \eqref{con_EM}. As we know, this provides us with a statement about the conservation of energy. By expanding the zeroth component, we have 
\begin{align}\label{eq_E}
&\frac{1}{c}\partial_t\big(\sqrt{-g}T^{00}\big)+\partial_j\big(\sqrt{-g}T^{0j}\big)+\Gamma^{0}_{00}\big(\sqrt{-g}T^{00}\big)+2\Gamma^{0}_{0j}\big(\sqrt{-g}T^{0j}\big)+\Gamma^{0}_{jk}\big(\sqrt{-g}T^{jk}\big)=0,
\end{align}
each term of which should be calculated up to the 1\tiny PN \normalsize order. The relevant calculations are summarized in Appx. \ref{app1}.  
Gathering together the results \eqref{eq1}-\eqref{eq5}, we arrive at
\begin{align}\label{zeroth-com-cons}
\nonumber
&\partial_t\rho^*+\partial_j\big(\rho^*v^j\big)+\frac{1}{c^2}\bigg\lbrace\partial_t\Big[\rho^*\big(\frac{G\,M}{r}+\frac{v^2}{2}+\Pi\big)\Big]
+\partial_j\Big[\rho^*\,v^j\big(\frac{G\,M}{r}
+\frac{v^2}{2}+\Pi\big)\Big]+\partial_j\big(p\,v^j\big)\\&+\frac{G\,M}{r^2}\rho^*\Big[\partial_t r +2\partial_j r\, v^j\Big]\bigg\rbrace+O(c^{-4})=0,
\end{align}
after some simplification. 
Moreover, given Eq. \eqref{cont_eq}, one can simplify the above relation and obtain the following equation:
\begin{align}
&\rho^*\partial_t\Big(\frac{v^2}{2}+\Pi\Big)+\rho^*v^j\partial_j\Big(\frac{v^2}{2}+\Pi\Big)+\partial_j\big(p\,v^j\big)+\frac{G\,M}{r^2}\partial_jr\,\rho^*v^j=0.
\end{align}
It is the local conservation of the energy within the fluid.
It should be noted that to further simplify this equation, we also use the Newtonian Euler equation
\begin{align}\label{N-Euler_eq}
&\rho^*\frac{dv^j}{dt}=-\partial_jp-\frac{G\,M}{r^2}\partial_jr\,\rho^*,
\end{align}
as well as the definition $d/dt=\partial_t+v^k\partial_k$. In the following derivation, it is shown that Eq. \eqref{N-Euler_eq} is obtained from the spacial component of Eq. \eqref{con_EM} in the Newtonian order (the 0\tiny PN \normalsize order). Regarding these points, we finally get 
\begin{align}\label{eq_Pi}
\frac{d\Pi}{dt}=\frac{p}{{\rho^*}^2}\frac{d\rho^*}{dt}+O(c^{-2}).
\end{align}
This is in fact the first law of thermodynamics for perfect fluids. As expected, the spin effects do not appear in this equation up to the order kept here. It is worth noting that it is sufficient to know the energy equation to the 0\tiny PN \normalsize order as written in Eq. \eqref{eq_Pi}. This is because the terms including the internal energy will eventually appear with the coefficient $\frac{1}{c^2}$ in the hydrodynamic equations. Therefore, the $O(c^{-2})$ terms in Eq. \eqref{eq_Pi} play a role in the 2\tiny PN \normalsize corrections, namely $O(c^{-4})$. This point is clarified in the next paragraphs.

In a similar fashion, we attempt to obtain the spatial components of Eq. \eqref{con_EM}. This will be a statement of momentum conservation. By expanding this component, we obtain
\begin{align}\label{s_com}
&\frac{1}{c}\partial_t\big(\sqrt{-g}T^{0j}\big)+\partial_k\big(\sqrt{-g}T^{jk}\big)+\Gamma^{j}_{00}\big(\sqrt{-g}T^{00}\big)+2\Gamma^{j}_{0k}\big(\sqrt{-g}T^{0k}\big)+\Gamma^{j}_{kn}\big(\sqrt{-g}T^{kn}\big)=0. 
\end{align}
The five terms in the above relation are given in Eqs. \eqref{eq6}-\eqref{eq10} to the required order. 
Substituting them into Eq. \eqref{s_com}, one can find 
\begin{align}\label{eqEuler}
\nonumber
&\partial_t\big(\mu\rho^*v^j\big)+\partial_k\big(\mu\rho^*v^jv^k\big)+\partial_jp+\frac{G\,M\,\rho^*}{r^2}\partial_jr+\frac{G\,M\,\rho^*}{c^2r^2}
\partial_jr\Big[\Pi+\frac{p}{\rho^*}-\frac{3\,G\,M}{r}+\frac{3}{2}v^2\Big]\\\nonumber&+\frac{G\rho^*}{c^2r^2}\bigg\lbrace\frac{2}{r}\Big[\big(\bm{x}\times\partial_t\bm{S}\big)_j+\big(\partial_t\bm{x}\times\bm{S}\big)_j\Big]-\frac{6}{r^2}\partial_tr\big(\bm{x}\times\bm{S}\big)_j-2M\,v^j\big(\partial_t r+v^k\partial_kr\big)\\
&+2v^k\Big[\frac{3}{r^2}\Big(\partial_jr\big(\bm{x}\times\bm{S}\big)_k-\partial_kr\big(\bm{x}\times\bm{S}\big)_j\Big)-\frac{1}{r}\Big(\big(\partial_j\bm{x}\times\bm{S}\big)_k-\big(\partial_k\bm{x}\times\bm{S}\big)_j\Big)\Big]\bigg\rbrace+O(c^{-4})=0,
\end{align}
where $\mu \coloneqq 1+\frac{1}{c^2}\big(\Pi+\frac{p}{\rho^*}+\frac{G\,M}{r}+\frac{v^2}{2}\big)$. The sum of the first two terms in this relation can be written as follows:
\begin{align}
\partial_t\big(\mu\rho^*v^j\big)+\partial_k\big(\mu\rho^*v^jv^k\big)=\rho^*v^j\frac{d\mu}{dt}+\mu\,\rho^*\frac{dv^j}{dt}. 
\end{align}
To obtain this result, the continuity equation \eqref{cont_eq} and the definition of the total time derivative are utilized. Inserting this relation into Eq. \eqref{eqEuler} and then truncating the result to the Newtonian order yield the Newtonian Euler equation mentioned earlier in Eq. \eqref{N-Euler_eq}. This equation governs the behavior of the Newtonian fluid in a gravitational field far from a spinning body where only its Newtonian effects are significant. As can be seen, only the mass of the body is involved in this result, not its spin. So, we go one step further and examine the next post-Newtonian terms, \textit{i.e.,} the 1\tiny PN \normalsize order, in the equation of motion to investigate the possible relativistic spin as well as mass effects.  To do so, the total time derivative of $\mu$ should be derived. Regarding the definition of $\mu$, one can easily show that
\begin{align}
\frac{d\mu}{dt}=\frac{1}{c^2}\Big[\frac{1}{\rho^*}\partial_tp-\frac{2\,G\,M}{r^2}v^k\partial_kr-\frac{G\,M}{r^2}\partial_tr\Big].
\end{align}
Using this relation and doing some manipulations, we finally arrive at 
\begin{align}\label{Eul_PN}
\nonumber
&\rho^*\frac{dv^j}{dt}=-\partial_jp-\frac{G\,M\,\rho^*}{r^2}\partial_jr+\frac{1}{c^2}\bigg\lbrace\Big(\Pi+\frac{p}{\rho^*}+\frac{G\,M}{r}+\frac{v^2}{2}\Big)\times\partial_jp-v^j\partial_tp\bigg\rbrace \\\nonumber
&-\frac{G\,\rho^*}{c^2r^2}\bigg\lbrace M\Big(v^2-\frac{4\,G\,M}{r}\Big)\partial_jr-M\,v^j\big(3\,\partial_tr+4\,v^k\partial_kr\big)+\frac{2}{r}\Big[\big(\bm{x}\times\partial_t\bm{S}\big)_j+\big(\partial_t\bm{x}\times\bm{S}\big)_j\Big]\\\nonumber
&-\frac{6}{r^2}\partial_t r\big(\bm{x}\times\bm{S}\big)_j+2v^k\Big[\frac{3}{r^2}\Big(\partial_j r\big(\bm{x}\times\bm{S}\big)_k-\partial_k r\big(\bm{x}\times\bm{S}\big)_j\Big)-\frac{1}{r}\Big(\big(\partial_j\bm{x}\times\bm{S}\big)_k-\big(\partial_k\bm{x}\times\bm{S}\big)_j\Big)\Big]\bigg\rbrace\\
&+O(c^{-4}).
\end{align}
This relation is the Euler equation of a non-gravitational post-Newtonian fluid embedded in the post-Newtonian field of a spinning compact body. The terms within the braces are the relativistic corrections to the Newtonian Euler equation. It reveals that even down to the leading relativistic order, the 1\tiny PN \normalsize order, the governing equations are very complicated, and in several terms, the relativistic effects of the central body and those of the fluid system manifest themselves 
\footnote{According to our knowledge, this equation has not been presented in the literature in this way.}.
Removing the spin terms, the above relation reduces to the hydrodynamic equation describing a relativistic fluid in the gravitational field induced by the exterior geometry of a compact body with mass $M$ up to the 1\tiny PN \normalsize order. It is a reasonable result revealing we are on the right track.
  
As pointed out before, the internal energy comes as a post-Newtonian correction in the Euler equation (see the third term on the right-hand side of Eq. \eqref{Eul_PN}). It means Eq. \eqref{eq_Pi} will give us enough information and we do not need to consider higher-order corrections to the energy equation. It should also be noted that the above result is a general relation that can be used to describe general systems without imposing any restrictive assumptions, such as aligned fluid models, \textit{i.e.,} situating the fluid system in the equatorial plane of the central body, steady-state models, etc. 
To compare this result with those in GR, however, we restrict ourselves to some assumptions. 
Despite these limitations, it is seen that relativistic properties of the central body, especially the spin effects, still play a role in simplified hydrodynamic equations, which indeed allow us to take another step towards studying the critical points in the system.

Up to this point, five hydrodynamic equations, \textit{i.e.,} three equations from Eq. \eqref{Eul_PN} and two equations from Eqs. \eqref{eq_Pi} and \eqref{cont_eq}, are introduced to describe a perfect fluid with six unknowns $\big(\rho^*, p, \Pi, \bm{v}\big)$. 
As usual, the last equation complementing this set of hydrodynamic equations is the EoS. In hydrodynamics, a class of the equations of state is considered as barotropic, which states that pressure is only a function of the density $\rho$ (see \cite{2013rehy.book.....R} and references therein).  
On the other hand, in the context of post-Newtonian gravity, $\rho^*$ has been applied as the rescaled mass density everywhere. Then following this scheme, we consider that pressure is a function of $\rho^*$ to define a barotropic fluid in the post-Newtonian framework. In  \cite{poisson2014gravity,2017ApJ...839...75N,2018ApJ...865...71K}, it is also considered that $p(\rho^*)$ in the context of the post-Newtonian gravity. So the EoS is given by
\begin{align}\label{EOS}
p=p(\rho^*),
\end{align}
from which, one can define the post-Newtonian sound speed as 
\begin{align}
\big(C_{\rm s}^2\big)_{\rm PN}\coloneqq \Big(\frac{\partial p}{\partial \rho^*}\Big)_{{s}_0},
\end{align}
where ${s}_0$ is the specific entropy. In the following, after specifying the form of Eq. \eqref{EOS}, we derive the relation between post-Newtonian and Newtonian sound speeds.

\section{Post-Newtonian hydrodynamic equations}\label{sec4}
As it turns out, the post-Newtonian hydrodynamic equations describing a perfect fluid around spinning bodies are very complicated. To proceed further toward solving these relations and comparing the results with those in GR, we impose the common assumptions used in standard accretion fluid studies. We choose spherical polar coordinates $(r,\, \theta,\,\varphi)$, centered on the body.
In this coordinate system, $\bm{x}=r \,\hat{r}$ and  $\bm{v}=\dot{r}\,\hat{r}+r\,\dot{\theta}\,\hat{\theta}+r\,\dot{\varphi}\,\sin\theta\,\hat{\varphi}$ represent the field point and the velocity vector of the fluid element, respectively. Here, the overdot indicates $d/dt$.
We consider that the fluid system around the central body is a geometrically thin disk\footnote{We assume that compared to the radial structure of the disk plane ($r$), the local half-thickness of the disk, $H$, is much smaller \textit{i.e.,} $H/r < 1$.}
and is situated in the equatorial plane of the compact body, whose spin is aligned with the $z$-axis. Therefore, we set $\theta=\frac{\pi}{2}$\footnote{In other words, the system is \textit{a priori} assumed to have an axisymmetric configuration around the $z$-axis. Nevertheless, a misaligned disk whose normal vector is not parallel to the $z$-axis, can also be thoroughly examined utilizing the general equations presented in the current work. We leave this as a future study.}. Regarding that, the stream has no motion in the transverse direction, \textit{i.e.,} $\dot{\theta}=0$, and its angular momentum essentially aligns with the angular momentum of the body.

Furthermore, we assume that the spin vector of the body is constant with respect to time, and consequently, we set $d\bm{S}/dt=0$. On the other hand, according to the natural definition of spin, based on the macroscopic rotation of the compact body, we know that $\bm{S}$ is essentially a time-dependent vector. There are two types of effects with completely different natures that can cause spin evolutions. One of them is the Newtonian effect being a result of the interaction between the extended body's multipole moments and the external gravitational potentials. As the external potential, \textit{i.e.,} the self-gravity of the disk, is utterly negligible in our case study, this non-relativistic effect does not lead to any precession of spin. However, due to other effects appearing in the post-Newtonian order, spin may still experience a precession. These relativistic effects are shown to be induced by spin-orbit and spin-spin couplings, called the geodetic precession and frame-dragging precession (Lense-Thirring precession), respectively. For details, see \cite{1993tegp.book.....W,poisson2014gravity}. Of course, for the system considered here in which only the central body has spin, only the spin-orbit precession can occur. So, basically, we have $d\bm{S}/dt=O(c^{-2})$. Keeping this fact in mind and regarding that all spin terms in the equations of motion \eqref{Eul_PN} appear as post-Newtonian corrections, the mentioned relativistic effects are indeed the 2\tiny PN \normalsize corrections. Therefore, we continue our study with the reasonable assumption $d\bm{S}/dt=0$ and neglect the spin evolution in the current work.

\subsection{Model assumptions}

Applying the above assumptions, we obtain the hydrodynamic equations governing the accretion dynamics around the spinning compact body including the 1\tiny PN \normalsize corrections related to the spin and mass of the body as well as the relativistic effects of the fluid.  
After some calculations and manipulations, we arrive at 
\begin{align}\label{Euler_r}
\nonumber
&\partial_tv_r+v_r\partial_rv_r+\frac{v_{\varphi}}{r}\partial_{\varphi}v_r-\frac{v^2_{\varphi}}{r}=-\frac{1}{\rho^*}\partial_rp-\frac{G\,M}{r^2}+\frac{1}{c^2}\bigg\lbrace\frac{1}{\rho^*}\partial_rp\Big(\Pi+\frac{p}{\rho^*}+\frac{G\,M}{r}+\frac{1}{2}v^2\Big)\\ &-\frac{1}{\rho^*}v_r\partial_tp+\frac{G\,M}{r^2}\Big(\frac{4\,G\,M}{r}+3\,v_r^2-v_{\varphi}^2+\frac{2\,s\,v_{\varphi}}{r}\Big)\bigg\rbrace+O(c^{-4}),
\end{align}
and
\begin{align}\label{Euler_phi}
\nonumber &\partial_tv_{\varphi}+v_r\partial_rv_{\varphi}+\frac{v_{\varphi}}{r}\partial_{\varphi}v_{\varphi}+\frac{v_r\,v_{\varphi}}{r}=-\frac{1}{\rho^*\,r}\partial_{\varphi}p+\frac{1}{c^2}\bigg\lbrace\frac{1}{\rho^*\,r}\partial_{\varphi}p\Big(\Pi+\frac{p}{\rho^*}+\frac{G\,M}{r}+\frac{1}{2}v^2\Big)\\
&-\frac{1}{\rho^*}v_{\varphi}\partial_tp+\frac{2\,G\,M\,v_r}{r^2}\Big(2\,v_{\varphi}-\frac{s}{r}\Big)\bigg\rbrace+O(c^{-4}),
\end{align}
for the radial and azimuthal components of the post-Newtonian Euler equation \eqref{Eul_PN}, respectively. 
Here, $v_r\coloneqq\dot{r}$ is the radial \textit{drift} velocity, $v_{\varphi}\coloneqq r\dot{\varphi}$ is the circular/azimuthal velocity, and $s\coloneqq\frac{S}{M}$ is the spin of the body per unit mass. 

These results reveal that the spin affects both the radial and azimuthal components. In other words, a spin-orbit acceleration is added to the Euler equations as the 1\tiny PN\normalsize-order term\footnote{It should be noted that although in the equations of motion, the spin-orbit effects are formally of the 1PN order, for the compact body, they are effectively of the 1.5PN order. However, similar to the first post-Newtonian description of the $N$-body system, we retain these terms in the equations of motion \eqref{Euler_r} and \eqref{Euler_phi}.}, see the last terms of Eqs. \eqref{Euler_r} and \eqref{Euler_phi}.  
This is indeed a relativistic interaction between the body's spin and the motion of fluid elements inside the gravitational potential of the body. In the post-Newtonian description of the $N$-body system, it is shown that spin-orbit accelerations, as well as spin-spin accelerations, play a role in the equations of motion \cite{poisson2014gravity,1993tegp.book.....W}. Here, since the central body is the only object with spin, we do not encounter acceleration due to the spin-spin effect.     
Moreover, in the following where these equations are obtained in the fluid element's comoving frame, we will find similar spin-orbit corrections that originate from the fourth term on the left-hand side of Eqs. \eqref{Euler_r} and \eqref{Euler_phi}, \textit{i.e.,} $v^2_{\varphi}/r$ and $v_r\,v_{\varphi}/r$, that are in principle related to the $r$- and $\varphi$-directed Coriolis forces due to frame dragging induced by the central rotating body. 
Furthermore, during the following calculations, where we attempt to further simplify the hydrodynamic equations, we will encounter footprints of the coupling of the body spin with itself, which is bilinear in $S$. This so-called spin-spin effect is in fact of the order $c^{-4}$. It means that this type of coupling, whose underlying physics can be conceptually of interest, may play a role in hydrodynamics at the next post-Newtonian order, 2\tiny PN \normalsize order. 

Our next task is to derive the Euler equation in the vertical direction for the thin disk following the post-Newtonian geometry. By the vertical direction, we mean the direction orthogonal to the disk plane, \textit{e.g.}, the $z$ ($\theta$) direction in cylindrical (spherical) coordinates. Although our formulation is based on the spherical coordinate system, in the following, for the sake of simplicity, we obtain the vertical equation in cylindrical coordinates. Of course, the final physical result will be independent of the coordinates chosen. 
Like the previous two equations, we obtain the $z$-component of the post-Newtonian Euler equation \eqref{Eul_PN} as

\begin{align}\label{Euler_z}
\nonumber
&\rho^*\ddot{z}=-\partial_z p-\frac{G\,M\,\rho^*}{r^3}z+\frac{1}{c^2}\bigg\lbrace\partial_z p\Big(\Pi+\frac{p}{\rho^*}+\frac{G\,M}{r}+\frac{1}{2}v^2\Big)\\
&+\frac{G\,M\,\rho^*}{r^3}z\Big(\frac{4\,G\,M}{r}-v^2+\frac{6\,s\,v_{\varphi}}{r}\Big)\bigg\rbrace+O(c^{-4}).
\end{align}
Here, $r=\sqrt{R^2+z^2}$ where $R$ is the radial distance from the $z$-axis and $z$ is the
height above the equatorial plane. In a thin-disk approximation where at the radius $R$, the thickness of the disk $2H$ is always much less than $R$, we have $r\simeq R$. 
It can be seen that spin-orbit acceleration appears in this relation as well. Also, based on the assumption that the angular momentum of the body and the disk are parallel, it is natural to expect that the general relativistic Coriolis force, caused by frame dragging, has no component in the $z$-direction. Therefore, this type of force plays no role in the vertical Euler equation, as it does in Eq. \eqref{Euler_z}.

The equations \eqref{Euler_r}, \eqref{Euler_phi}, and \eqref{Euler_z} capture the basic characteristics of fluid motion around a spinning compact body in the post-Newtonian gravity. It should be noted that the terms within the braces in these relations are the post-Newtonian corrections to the Newtonian equations.

\subsection{Accretion flows in steady-state}

In this section, we focus on advection dominated, axisymmetric accretion flows  in a steady state. To describe a steady-state, we need to consider $ \partial/\partial t=0$ and for the axisymmetry $\partial/\partial \varphi=0$. In the relativistic framework, \cite{1973blho.conf..343N} have introduced steady-state models for the accretion disk around a Kerr black hole. In the current work, we shall introduce the steady-state post-Newtonian models for the accretion disk embedded in the post-Newtonian gravitational field of a spinning body. We also follow the standard approximation in accretion flows where the mean flow is vertically averaged \cite{1998ApJ...498..313G}. The vertical averaging approximation of a flow variable $f$ is introduced as
\begin{align}
\int f d\Omega \simeq 4\pi H_{\theta} f_0,
\end{align}
where $d\Omega\coloneqq \sin \theta d\theta d\varphi$  is an element of solid angle and $f_0$ is the flow variable in the disk plane, $f_0=f(\theta=\pi/2)$. The integration interval is from $\varphi=0$ to $\varphi=2\pi$, and from $\theta=\pi/2-H_{\theta}$ to $\theta=\pi/2+H_{\theta}$.
Here, $H_{\theta}$ is the characteristic angular scale of the flow about the equator. In the thin disk approximation, the local half-thickness of the disk $H$ in terms of this scale can be considered as $H\simeq H_{\theta} r$. Moreover, as another common assumption in the semi-analytic accretion flows study, we consider that the fluid system is in hydrostatic vertical equilibrium, namely $\ddot{z}=0$ in Eq. \eqref{Euler_z}.

To analyze the physical processes around the spinning bodies, we finally derive the dynamical equations in the local rest frame (LRF) of the fluid. In order to transform from the global frame to LRF and vice versa, we follow the standard procedure applied by \cite{1998ApJ...498..313G}. In a similar fashion, we introduce two important frames: $1-$ locally nonrotating frame (LNRF). $2-$ corotating frame (CRF). 
The LNRF is attached to an observe whose worldline is $\theta=const.$, $r=const.$, and $\varphi=\omega\,t+const.$, where $\omega$ is the rate of frame dragging by the central spinning body and is obtained in terms of the spin in Eq. \eqref{omega}. In the paper of \cite{1972ApJ...178..347B}, the transformations between a global coordinate frame in a general, stationary, axisymmetric, asymptotically flat spacetime and the LNRF have been introduced. 
The next frame corotating with the fluid is called CRF. This frame moves with respect to the LNRF with a constant velocity $\beta_{\varphi}$ confined to the $\varphi$-direction \footnote{Here, we utilize the notation used by \cite{1998ApJ...498..313G}.}. Therefore, using the Lorentz transformations with the Lorentz factor $\gamma_{\varphi}= \big(1-\beta^2_{\varphi}/c^2\big)^{-1/2}$, it is easy to transform from the LNRF to the CRF.
On the other hand, the LRF has a constant radial velocity, which is hereafter denoted by $V$, with respect to CRF. Due to the accretion of the material onto the central body, $V$ is indeed negative. The LRF can also be obtained by a Lorentz boost with $\gamma_{r}= \big(1-V^2/c^2\big)^{-1/2}$ from the CRF. In this way, with the help of two auxiliary frames, \textit{i.e.,} LNRF and CRF, with an appropriate combination of coordinate transformations, one can reach the LRF from the global frame. In Appx. \ref{app2}, the relevant calculations are given in detail.      

The rest of this section is dedicated to introducing the post-Newtonian hydrodynamic equations in the LRF by imposing the aforementioned assumptions.

\subsubsection{Particle number conservation}

In the case of the steady axisymmetric model, after performing the vertical averaging, Eq. \eqref{cont_eq} reduces to  

\begin{align}
\partial_r\left(4\pi H_{\theta} r^2\Big(1+\frac{2\,G\,M}{c^2\,r}\Big)\rho\,u^r\right)=0,
\end{align}
where the flow variables are computed at $\theta=\pi/2$. Integrating over radius gives
\begin{align}
4\pi H r\rho u^r\Big(1+\frac{2\,G\,M}{c^2\,r}\Big)=-\dot{M}.
\end{align}
Here, $\dot{M}$ is a constant interpreted as the rest-mass accretion rate. Next,
using Eq. \eqref{u^mu} and writing $u^r$ in terms of $V$, we arrive at 
\begin{align}\label{Mdot-PN}
4\pi H r\rho V\Big[1+\frac{1}{c^2}\Big(\frac{G\,M}{r}+\frac{1}{2}V^2\Big)\Big]=-\dot{M}.
\end{align}
This relation renders a constant inflow of the rest mass passing through each radius of the post-Newtonian accretion disk \footnote{Of course, this statement is reliable only in the interval $r_{\text{PN}} \leqslant r \leqslant r_{\infty}$ which is the realm of validity of the current study.}. 

It is worth noting that although spin effects do not play a role here, two other post-Newtonian corrections related to the gravitational potential of the central body and the radial velocity of the fluid system appear in this relation.
In accordance with the nature of the accretion phenomenon, these relativistic effects must have a strengthening contribution to $\dot{M}$. The 1\tiny PN\normalsize-order result \eqref{Mdot-PN} is in beautiful agreement with this fact.  The faster the radial velocity and the stronger the body's gravity, the higher the rest-mass accretion rate. 
By removing these corrections, one reaches the standard form of the accretion rate $4\pi H r\rho V=-\dot{M}$ in the non-relativistic regime, \textit{e.g.}, cf. \cite{2002apa..book.....F,2008bhad.book.....K}.

\subsubsection{Radial momentum equation}
Starting from Eq. \eqref{Euler_r}, one easily deduces the radial component of the Euler equation as
\begin{align}\label{Euler_r_N_ss}
\nonumber
&v_r\partial_rv_r=\frac{v^2_{\varphi}}{r}-\frac{1}{\rho^*}\partial_rp-\frac{G\,M}{r^2}+\frac{1}{c^2}\bigg\lbrace\frac{1}{\rho^*}\partial_rp\Big(\Pi+\frac{p}{\rho^*}+\frac{G\,M}{r}+\frac{1}{2}v^2\Big)\\
&+\frac{G\,M}{r^2}\Big(\frac{4\,G\,M}{r}+3\,v_r^2-v_{\varphi}^2+\frac{2\,s\,v_{\varphi}}{r}\Big)\bigg\rbrace,
\end{align}
for the steady-axisymmetric model. The next task is to further simplify this radial momentum equation and obtain it in terms of the preferred variables $V$ and $\beta_{\varphi}$.

To do so, let us first derive the required ingredients $\gamma$, $v_{\varphi}$, and $v_{r}$. It is defined that $\gamma=\frac{u^0}{c}$. Then regarding the relation between $u^0$ and the preferred variables given in Eq. \eqref{u^mu}, we get
\begin{align}
\gamma=\gamma_{\text{tot}}\Big(1+\frac{1}{c^2}\frac{G\,M}{r}-\frac{1}{c^4}\frac{G^2M^2}{r^2}\Big),
\end{align}
where $\gamma_{\text{tot}}=\gamma_r\gamma_\varphi$ is the total Lorentz factor. 
So, we have 
\begin{align}\label{gammaV}
\gamma=1+\frac{1}{c^2}\Big(\frac{G\,M}{r}+\frac{1}{2}V^2+\frac{1}{2}\beta_{\varphi}^2\Big),
\end{align}
after substituting the definition of $\gamma_{\text{tot}}$ and truncating the result to the 1\tiny PN \normalsize order. Comparing Eq. \eqref{gamma} with the above result reveals that $v^2\simeq V^2+\beta_{\varphi}^2$.  This relation can be used to determine $\rho^*$ in terms of $V$ and $\beta_{\varphi}$. 
In our notation, the radial component of the four-velocity is defined as $u^r=\gamma \dot{r}=\gamma v_r$. Therefore, considering Eq. \eqref{u^mu} for $u^r$ as well as Eq. \eqref{gammaV} for $\gamma$, we then have 
\begin{align}\label{v_r}
v_r=V\Big[1-\frac{1}{c^2}\Big(\frac{2\,G\,M}{r}+\frac{1}{2}\beta_{\varphi}^2\Big)\Big]. 
\end{align}
In a similar fashion, using $u^\varphi=\gamma \dot{\varphi}$, we obtain 
\begin{align}\label{v_varphi}
v_{\varphi}=\frac{\ell}{r}\Big[1-\frac{1}{c^2}\Big(\frac{3\,G\,M}{r}+\frac{1}{2}V^2+\frac{1}{2}\beta_{\varphi}^2\Big)\Big]+\frac{G\,M\,s}{c^2r^2},
\end{align}
where $\ell$ is the angular momentum per unit mass of the fluid element, \textit{i.e.,} the specific angular momentum. Notice that $v_{\varphi}=r\dot{\varphi}$. The last term in Eq. \eqref{v_varphi} corresponds to the drag of the fluid element along the rotation of the central body. This term contains information about the Coriolis force.  Regarding the relation between $\ell$ and $\beta_\varphi$, given in Eq. \eqref{beta_phi}, it is obvious that $\ell$ is indeed a relativistic quantity.  

With Eqs. \eqref{gammaV}-\eqref{v_varphi} and the definitions \eqref{rho*PN} and \eqref{beta_phi}, after some simplifications, the radial component of the post-Newtonian Euler equation \eqref{Euler_r_N_ss} becomes
\begin{align}\label{Euler_V}
\nonumber
& V\partial_r V\Big[1-\frac{1}{c^2}\Big(\frac{4\,G\,M}{r}+\frac{\ell^2}{r^2}\Big)\Big]=\frac{\ell^2}{r^3}-\frac{1}{\rho}\partial_rp-\frac{G\,M}{r^2}-\frac{1}{c^2}\bigg\lbrace\frac{\ell^2}{r^3}\Big(3\,V^2+\frac{\ell^2}{r^2}\Big)-2\frac{\ell\,V^2}{r^2}\partial_r\ell\\
&-\frac{1}{\rho}\partial_r p\Big(\Pi+\frac{p}{\rho}+\frac{4\,G\,M}{r}+V^2+\frac{\ell^2}{r^2}\Big)-\frac{G\,M}{r^2}\Big(\frac{4\,G\,M}{r}+V^2-\frac{7\,\ell^2}{r^2}+\frac{4\,s\,\ell}{r^2}\Big)\bigg\rbrace.
\end{align}

It is constructive to mention that, as brought up earlier, the spin-spin effect interestingly manifests itself in the equations of motion. According to the relation \eqref{v_varphi}, at the lowest order, this kind of acceleration/interaction is hidden in the first and last terms on the right-hand side of Eq. \eqref{Euler_r_N_ss}. Notice that this effect would be like a \textit{self}-spin interaction and is bilinear in the body spin. Clearly, it can exist in the 2\tiny PN \normalsize order, but definitely not in the 1\tiny PN \normalsize order. 
We should keep in mind that in order to have a complete description of spin-spin effects, in addition to the aforementioned terms, the rest of the $O(c^{-4})$ terms that were omitted from the Eq. \eqref{Euler_r_N_ss} in the beginning, must be brought back into play. A full investigation of these effects is beyond the scope of the current study. 

\subsubsection{Azimuthal angular momentum equation}
For the steady axisymmetric model, the azimuthal component of the Euler equation, \textit{i.e.,} Eq. \eqref{Euler_phi}, can be written as 
\begin{align}
\label{Euler_phi_N_ss}
& \partial_rv_{\varphi}+\frac{v_{\varphi}}{r}=-\frac{1}{c^2}\frac{G\,M}{r^2}\Big(\frac{2\,s}{r}-4\,v_{\varphi}\Big),
\end{align}
which reduces to 
\begin{align}\label{ell-PN}
\partial_r\ell=-\frac{1}{c^2}\Big[\frac{G\,M}{r^2}\big(s-\ell\big)+\frac{\ell^3}{r^3}-\ell\,V\partial_r V\Big],
\end{align}
considering Eq. \eqref{v_varphi} and expanding the result to $O(c^{-2})$. This relation indicates that the equation governing the relativistic angular momentum $\ell$ is complicated and a combination of the post-Newtonian corrections prevents it from being a constant. In other words, unlike the Newtonian case, the angular momentum of the rotating fluid can change with respect to the radius. We should emphasize that this change is triggered by relativistic/post-Newtonian effects.

To illustrate the gravitational effect of spin on the motion of a fluid element, let us provide a crude estimate. Keeping only the Newtonian and spin terms in the relation above, we have 
\begin{align}
\partial_r\ell=-\frac{1}{c^2}\frac{G\,M}{r^2}s,
\end{align}
and integration yields
\begin{align}\label{ell_app}
\ell_{\infty}&=\ell-\frac{1}{c^2}\frac{G\,M\,s}{r},
\end{align}
where $\ell_{\infty}$ is an integration constant interpreted as the specific angular momentum of the fluid element at infinity. Here, $r$ lies in the interval $r_{\text{PN}} \leqslant r \leqslant r_{\infty}$. This relation reveals the consequence of frame dragging related to the 1\tiny PN\normalsize-order spin effect of the central body in the angular momentum of the fluid. To show this fact, we consider a possible case. A particle whose angular momentum vanishes at infinity, \textit{i.e.,} $\ell_{\infty}=0$, falls onto a spinning body. Regarding Eq. \eqref{ell_app}, it will eventually acquire an angular momentum equal to $G\,M\,s/(c^2r)$ at radius $r$, being in the same direction as the angular momentum of the body. In fact, the rotation of the central body drags the particle. This is a gravitational influence of spin.       
At the level of Lagrangian description, in the framework of post-Newtonian gravity, this point is mentioned in \cite{poisson2014gravity}.

\subsubsection{Internal energy and equation of state}\label{Internal_Energy}
In the case of the steady and axisymmetric disk, the energy equation \eqref{eq_Pi} reduces to 
\begin{align}
\partial_r\Pi =\frac{p}{{\rho^*}^2}\partial_r\rho^*.
\end{align}
As calculated, $\Pi$ appears in the equations of motion as a post-Newtonian term. Keeping this fact in mind, we can simply convert $\rho^*$ to $\rho$ in the above relation and remove the $O(c^{-2})$ terms. We only need to take them into account whenever we investigate the fluid behavior with higher relativistic corrections, at least up to the 2\tiny PN \normalsize order. Therefore, here, the required energy equation is given by 
\begin{align}\label{Pi}
\partial_r\Pi =\frac{p}{{\rho}^2}\partial_r\rho.
\end{align}

Now, to obtain accretion solutions, one requires to use the EoS, which describes a relation among the thermodynamical variables, namely density ($\rho$), pressure ($p$), and internal energy ($\Pi$). Typically, the temperature of an accreting matter increases beyond $> 10^{10}$ K as the flow reaches within a few tens of Schwarzschild radius, and the relativistic EoS would be an apt choice, see \cite{2018PhRvD..98h3004D,2022MNRAS.516.5092M} and references therein. However, the effect of PN corrections is valid only in regions outside $\sim 10\,r_g$, so we stick to the simplest ideal EoS (IEoS) and consider that the fluid element undergoes reversible adiabatic changes. Regarding Eq. \eqref{EOS}, we then have: 
\begin{align}\label{EOS2}
p=K {\rho_*}^{\Gamma},
\end{align}
where $K$ is a constant of proportionality which is a measure of entropy, and $\Gamma$ is the adiabatic index that should be chosen less than $4/3$. Hence, we fix $\Gamma=13/9$ throughout to maintain the trans-relativistic nature (see Fig. 1f in \cite{2022MNRAS.516.5092M}) of the flow unless stated otherwise.
Also, for the ideal fluid, the internal energy per unit mass is given by (see for instance \cite{2023MNRAS.523.4431M}),
\begin{equation}\label{Pi-ideal}
\Pi = \frac{1}{(\Gamma-1)} \frac{p}{\rho}.
\end{equation}

As the last point of this part, let us define the post-Newtonian sound speed and its relation with the Newtonian one. Simply inserting Eq. \eqref{EOS2} into Eq. \eqref{EOS}, using the definition \eqref{rho*PN}, and finally expanding the result in terms of $c$, we obtain 
\begin{align}
\big(C_{\rm s}^2\big)_{\rm PN}=C_{\rm s}^2\Big[1+\frac{\Gamma-1}{c^2}\big(\frac{3\,G\,M}{r}+\frac{1}{2}v^2\big)\Big],
\end{align} 
in which $C_{\rm s}^2=K\,\Gamma\,\rho^{\Gamma-1}$ is the Newtonian sound speed.

\subsubsection{Vertical equilibrium in accretion flows}
As mentioned previously, we assume that the hydrostatic balance holds in the thin accretion disk around a spinning body in the $z$ direction, \textit{i.e.,} $\ddot{z}=0$ \cite{1973blho.conf..343N,1995ApJ...450..508R}. 
So, the vertical momentum equation \eqref{Euler_z} is expressed as 
\begin{align}
\nonumber
&\partial_z p\Big[1-\frac{1}{c^2}\Big(\Pi+\frac{p}{\rho^*}+\frac{G\,M}{r}+\frac{1}{2}v^2\Big)\Big]=-\frac{G\,M\,\rho^*}{r^3}z
\Big[1-\frac{1}{c^2}\Big(\frac{4\,G\,M}{r}-v^2+\frac{6\,s\,v_{\varphi}}{r}\Big)\Big].
\end{align}
Using Eqs. \eqref{rho*PN}, \eqref{v_r}, and \eqref{v_varphi} as well as expanding the result, we arrive at the post-Newtonian version of the hydrostatic equilibrium 
\begin{align}\label{hyd_equi}
\partial_z p\Big[1-\frac{1}{c^2}\Big(\Pi+\frac{p}{\rho}+\frac{4\,G\,M}{r}+V^2+\frac{\ell^2}{r^2}\Big)\Big]=\rho g_z^{\text{\tiny PN}},
\end{align}
where 
\begin{align}
g_z^{\text{\tiny PN}}=-\frac{G\,M}{r^3}z\Big[1-\frac{1}{c^2}\Big(\frac{4\,G\,M}{r}- V^2- \frac{\ell^2}{r^2}+\frac{6\,s\,\ell}{r^2}\Big)\Big], 
\end{align}
is the vertical component of the gravitational acceleration at the height $z$. 
Eq. \eqref{hyd_equi} governs the balance between the vertical gradient of pressure and the vertical gravitational force with the relevant post-Newtonian corrections.

To sum up, in the reference frame comoving with the fluid element, Eqs. \eqref{Mdot-PN}, \eqref{Euler_V}, \eqref{ell-PN}, \eqref{Pi-ideal}, and \eqref{hyd_equi} together with the EoS \eqref{EOS2} construct a complete set of differential equations to describe the behavior of a relativistic, steady thin accretion disk in the post-Newtonian spacetime of a spinning compact body.

\section{Transonic accretion solutions in Post-Newtonian space-time}\label{sec5}

To examine the importance of the gravitational relativistic effects and to evaluate the physics behind them, we classify the post-Newtonian equations in three specific cases: 1-- semi-relativistic (SR) flow. 2-- semi-Newtonian (SN) flow. 3-- non-relativistic (NR) flow, and evaluate the critical point conditions for them.  
Regarding the complexity of the relations introduced in the previous section, for the time we proceed with these three cases. We will study the full post-Newtonian case in the future.

\subsection{Semi-relativistic fluid}

It is shown that in the region $r>8\,r_g$, the ratio of the radial velocity of the fluid element to the speed of light is of the order $10^{-1}$, cf. \cite{2018MNRAS.473.2415S} and references therein. So, we have $\frac{V^2}{c^2}\sim10^{-2}$. Knowing this point, we consider that
\begin{align}\label{condi_Semi_R}
\frac{V^2}{c^2}\sim\epsilon,
\end{align}
where $\epsilon\rightarrow 0$. On the other hand, in the standard accretion disk model, it is expected that $V\ll \beta_{\varphi}$. So, we preserve the relativistic effects of azimuthal velocity compared to those of the radial one. In fact, it is assumed that in the radial direction, the fluid velocity does not exceed the non-relativistic limit, while in the azimuthal, it does. We call this system a semi-relativistic (SR) fluid and keep the correction $\beta^2_{\varphi}/c^2$, \textit{i.e.,} $\ell^2/(c^2r^2)$, related to the relativistic motion of the fluid in the hydrodynamic equations. 

Based on the condition \eqref{condi_Semi_R}, we arrive at 
\begin{equation}\label{Euler_V_SR}
 V\partial_rV=\frac{\ell^2}{r^3}-\frac{1}{\rho}\partial_r p-\frac{G\,M}{r^2}+\frac{1}{c^2}\bigg\lbrace\frac{1}{\rho}\partial_r p\Big(\Pi+\frac{p}{\rho}\Big)-\frac{4\,G\,M}{r^2}\Big(\frac{\ell^2}{r^2}-\frac{s\,\ell}{r^2}\Big)\bigg\rbrace,  
\end{equation}
for the radial component, and further introduce the effective potential for the SR case, $\Phi_{\rm SR}$. In analogy with Newtonian mechanics, the sum of terms on the right-hand side of Eq. \eqref{Euler_V_SR}, except those related to the pressure force, is considered an effective force. So, after integrating under a fixed $\ell$, one can obtain 
\begin{align}\label{phi-PN}
\Phi_{\rm SR}=\frac{\ell^2}{2\,r^2}-\frac{G\,M}{r}\Big[1+\frac{4}{3\,c^2}\Big(\frac{\ell^2}{r^2}-\frac{s\,\ell}{r^2}\Big)\Big],
\end{align}
for the post-Newtonian effective potential,
\begin{align}\label{ell-SR}
\partial_r\ell=-\frac{1}{c^2}\Big[\frac{G\,M}{r^2}\big(s-\ell\big)+\frac{\ell^3}{r^3}\Big],
\end{align}
for the azimuthal component, and finally
\begin{align}\label{hyd_equi-SR}
\partial_z p=\rho g_z^{\text{\tiny SR}},
\end{align}
for the vertical component of the Euler equation. Here, 
\begin{align}\label{gz}
g_z^{\text{\tiny SR}}=-\frac{G\,M}{r^3}z\Big[1+\frac{1}{c^2}\Big(\Pi+\frac{p}{\rho}+ \frac{2\,\ell^2}{r^2}-\frac{6\,s\,\ell}{r^2}\Big)\Big]. 
\end{align}
Using Eqs. \eqref{hyd_equi-SR} and \eqref{gz}, one can show that the half-thickness of the SR disk is given by
 \begin{equation}\label{HSR}
 H_{\rm SR}= \frac{1}{\Omega_{\rm K}}\sqrt{\frac{p}{\rho}}\Big[1-\frac{1}{2\,c^2}\Big(\Pi+\frac{p}{\rho}+\frac{2\,\ell^2}{r^2}-\frac{6\,s\,\ell}{r^2}\Big)\Big],
 \end{equation}
where $\Omega_{\rm {K}}=\sqrt{\frac{G\,M}{r^3}}$ is the standard Keplerian angular velocity.  
To derive the above relation, we follow the scheme introduced by \cite{2008bhad.book.....K} and use the fact that pressure is zero on the surface of the disk.
Moreover, imposing the condition \eqref{condi_Semi_R} on Eq. \eqref{Mdot-PN} reveals that
\begin{align}\label{Mdot-SR}
4\pi H_{\text{\tiny SR}}\,r\rho V\Big[1+\frac{1}{c^2}\frac{G\,M}{r}\Big]=-\dot{M}_{\text{\tiny SR}},
\end{align}
in which $H_{\text{\tiny SR}}$ is described by Eq. \eqref{HSR}.  
As the final point, it should be noted that in this case, the post-Newtonian EoS \eqref{EOS2} simplified as 
\begin{align}\label{p-PN}
p=C_{\rm s}^2\frac{\rho}{\Gamma}\Big[1+\frac{\Gamma}{c^2}\Big(\frac{3\,G\,M}{r}+\frac{\ell^2}{2\,r^2}\Big)\Big],
\end{align}
and the internal energy in Eq. \eqref{Pi-ideal} are applied.

To solve for the feasible accretion solutions, we need to evaluate the dynamical equations corresponding to the flow velocity $V$, sound speed $C_{\rm s}$, and the specific angular momentum of the flow $\ell$, respectively. The latter one is already obtained in Eq. \eqref{ell-SR}. Further, with some algebraic steps involving equation of state (EoS) \eqref{Pi}-\eqref{EOS2}, and radial momentum equation \eqref{Euler_V_SR}, angular momentum equation \eqref{ell-SR}, one can substitute the mass-accretion rate Eq. \eqref{Mdot-SR} and obtain the radial velocity gradient (or wind equation) as
\begin{align}\label{wind-SR}
\frac{dV}{dr}|_{\rm SR} = \frac{N_{\rm SR} (r,V,C_{\rm s},\ell,s)}{D_{\rm SR}(r,V,C_{\rm s},\ell,s)},
\end{align}
and additionally, the temperature gradient equation as
\begin{align}\label{temp-grad-SR}
\frac{dC_{\rm s}}{dr} = C_{0_{\rm SR}} + C_{V_{\rm SR}} \frac{dV}{dr}.
\end{align}
The explicit form of $N_{\rm SR}$, $D_{\rm SR}$, $C_{0_{\rm SR}}$, and $C_{V_{\rm SR}}$ are given in Appx. \ref{SR_Wind}. It is a standard way to express the wind equation in terms of the ratio of numerator ($N_{\rm SR}$) and denominator ($D_{\rm SR}$), where they depend explicitly on the flow variables, namely, $r$, $V$, $C_{\rm s}$, $\ell$ and $s$. See \cite{2018PhRvD..98h3004D,2022MNRAS.516.5092M} and references therein. In the next Sec. \ref{transonic-method}, we will discuss obtaining global accretion solutions from these equations. 

\subsection{Semi-Newtonian fluid}\label{Semi_Newtonian_fluid}

In this case, it is assumed that the pressure and internal energy of the fluid are not relativistic. So, in addition to the radial velocity, these quantities do not exceed the non-relativistic limit.    
We then impose the following conditions  
\begin{align}\label{condi_Semi_N}
\frac{p}{\rho c^2}\sim \frac{\Pi}{ c^2}\sim\frac{V^2}{c^2}\sim\epsilon,
\end{align}
where $\epsilon\rightarrow0$. This is the semi-Newtonian (SN) approximation and the system that satisfies these conditions is called the SN system. This is because the matter distribution is still allowed to move rapidly in the $\varphi$ direction. The hydrodynamic equations governing the SN fluid are given below. 

The first one describes the radial structure as
\begin{align}\label{Euler_V_SN}
& V\partial_rV=\frac{\ell^2}{r^3}-\frac{1}{\rho}\partial_r p-\frac{G\,M}{r^2}\Big[1+\frac{4}{c^2}\Big(\frac{\ell^2}{r^2}-\frac{s\,\ell}{r^2}\Big)\Big].
\end{align}
The second one is the azimuthal Euler equation. We find that this relation is not affected by the SN conditions \eqref{condi_Semi_N} and Eq. \eqref{ell-SR} is recovered here. 
Moreover, in this case, for the vertical structure, we get
\begin{align}
\partial_z p=\rho\, g_z^{\text{\tiny SN}},
\end{align}
where
\begin{align}
g_z^{\text{\tiny SN}}=-\frac{G\,M}{r^3}z\Big[1+\frac{2}{c^2}\Big(\frac{\ell^2}{r^2}-\frac{3\,s\,\ell}{r^2}\Big)\Big].
 \end{align}
So, the half-thickness of the SN disk will be
\begin{align}
H_{\text{\tiny SN}}= \frac{1}{\Omega_{\text{K}}}\sqrt{\frac{p}{\rho}}\Big[1-\frac{1}{c^2}\Big(\frac{\ell^2}{r^2}-\frac{3\,s\,\ell}{r^2}\Big)\Big].
\end{align} 
Also, for this system, the mathematical form of the rest mass conservation is similar to the previous case, \textit{i.e.,} Eq. \eqref{Mdot-SR}, but the form of $H$ is different and is given by the above result. Then, we have
\begin{align}\label{Mdot-PN-SN}
4\pi H_{\text{\tiny SN}}\,r\rho V\Big[1+\frac{1}{c^2}\frac{G\,M}{r}\Big]=-\dot{M}_{\text{\tiny SN}}.
\end{align} 
Since no trace of internal energy has appeared in the calculated relations, we do not need Eq. \eqref{Pi-ideal} to describe the behavior of the SN fluid. In this case, the EoS is the same as in the previous case. 
Comparing Eqs. \eqref{Euler_V_SR} and \eqref{Euler_V_SN}, one can show that the effective potential in this case is equal to the previous case, $\Phi_{\rm SN}\equiv\Phi_{\rm SR}$ and it is described by Eq. \eqref{phi-PN}. 

Further, we follow Eqs. \eqref{Pi}, \eqref{EOS2}, \eqref{ell-SR}, \eqref{Euler_V_SN}, and \eqref{Mdot-PN-SN} to obtain the wind equation and the temperature gradient as in the previous case,
\begin{align}\label{wind}
\frac{dV}{dr}|_{\rm SN} = \frac{N_{\rm SN} (r,V,C_{\rm s},\ell,s)}{D_{\rm SN}(r,V,C_{\rm s},\ell,s)},
\end{align}
and the temperature gradient equation as
\begin{align}\label{temp-grad}
\frac{dC_{\rm s}}{dr} = C_{0_{\rm SN}} + C_{V_{\rm SN}} \frac{dV}{dr}.
\end{align}
The explicit form of $N_{\rm SN},D_{\rm SN},C_{0_{\rm SN}},C_{V_{\rm SN}}$ are given in Appx. \ref{SN_Wind}.

\subsection{Non-relativistic/Newtonian fluid}\label{Newtonian_fluid}
For this case, we ignore the post-Newtonian corrections 
\begin{align}\label{condi-N-1}
\frac{p}{\rho c^2}\sim \frac{\Pi}{ c^2}\sim\frac{V^2}{c^2}\sim\frac{\beta_{\varphi}^2}{c^2}\sim \epsilon,
\end{align}
which are related to the disk, setting $\epsilon\rightarrow 0$.  Relied heavily on this assumption, we consider that the fluid system is Newtonian/non-relativistic (NR). On the other hand, we keep post-Newtonian corrections associated with the central body. 
So, this NR system is indeed embedded in the background post-Newtonian spacetime. 
According to Eq. \eqref{beta_phi}, the term proportional to $\frac{\ell^2}{r^2 c^2}$ is what is omitted here. 
There are important consequences under this condition that must be considered during the calculation in this case.

We know that in the classical picture of the standard accretion disk \cite{1973A&A....24..337S}, in the absence of pressure as well as friction (viscosity), the centrifugal force is balanced by the gravitational force. So, for a Keplerian fluid around a body with mass M, we have 
\begin{align}
\ell^2_{\text{K}}=G\,M\,r.
\end{align}
On the other hand, the post-Newtonian corrections related to the central body are taken into account, and we keep terms like $\frac{G\,M}{c^2r}$ in the equations of motion. Keeping this point in mind, assuming a very small value for the ratio $\frac{\ell^2}{r^2 c^2}$ basically means forcing the fluid system to be sub-Keplerian throughout the disk, \textit{i.e.,} 
\begin{align}\label{condi-N-2}
\ell(r)\ll\ell_{\text{K}}(r).
\end{align}
Furthermore, in the post-Newtonian description, we consider that the spacetime deviates slightly from the flat spacetime and apply $\frac{G\,M}{c^2 r}\ll 1$. This condition dictates that $\ell^2_{\text{K}}\ll c^2r^2$. This point reveals that in the specific case studied here, even for an extreme Kerr black hole with $s=\frac{G\, M}{c}$, the post-Newtonian term  $\frac{s\,\ell}{c^2r^2}$ related to the gravitational/general relativistic effect of the black hole spin, is much smaller than the term $\frac{G\,M}{c^2r}$ related the general relativistic effect of the black hole mass. Therefore, in the following calculations, in addition to the mentioned terms, we consider that
\begin{align}\label{condi-N-3}
\frac{s\,\ell}{c^2r^2}\sim \epsilon,
\end{align}
where $\epsilon\rightarrow 0$.

Applying the above restrictions, we obtain the NR equations of motion. The first one is the radial structure which reduces to the well-known Newtonian case
\begin{align}\label{Euler-N}
& V\partial_r V=\frac{\ell^2}{r^3}
-\frac{1}{\rho}\partial_r p-\frac{G\,M}{r^2}.
\end{align} 
Therefore, we recover $\Phi_{\rm eff}^{\rm N}$ which is given in Eq. \eqref{Phi_eff_N}.
For the angular momentum equation \eqref{ell-PN}, imposing the conditions \eqref{condi-N-1}, \eqref{condi-N-2}, and \eqref{condi-N-3}, we find that for the NR case, $\ell$ is a constant. 
Moreover, the vertical structure and the rest mass conservation are described by
\begin{align}\label{pz}
\partial_z p=-\rho\frac{G\,M}{r^3}z,
\end{align}
and
\begin{align}\label{Mdot-N}
4\pi H_{\text{\tiny NR}}\rho rV\Big(1+\frac{G\,M}{c^2\,r}\Big)=-\dot{M}_{\text{\tiny NR}},
\end{align}
respectively. Here, $H_{\text{\tiny NR}}$ is the half-thickness of the NR disk, which is obtained from Eq. \eqref{pz} as follows:
\begin{align}
H_{\text{\tiny NR}}=\frac{1}{\Omega_{\text{K}}}\sqrt{\frac{p}{\rho}}.
\end{align} 
Further, the EoS is expressed as
\begin{align} 
p=C_{\rm s}^2\frac{\rho}{\Gamma}\Big[1+\frac{\Gamma}{c^2}\frac{3\,G\,M}{r}\Big].
\end{align}
Finally, we evaluate the wind equation as 
\begin{equation}
    \frac{dV}{dr}|_{\rm NR} = \frac{N_{\rm NR}(r,V,C_{\rm s},\ell,s)}{D_{\rm NR}(r,V,C_{\rm s},\ell,s)},
\end{equation}
and
\begin{equation}\label{temp-grad_SN}
    \frac{dC_{\rm s}}{dr} = C_{0_{\rm NR}} + C_{V_{\rm NR}} \frac{dV}{dr},
\end{equation}
where the coefficients are mentioned in detail in Appx. \ref{NR_Wind}.

Before introducing the accretion solutions in the post-Newtonian framework, let us discuss the effective potentials as the final point of this subsection. In order to express the flow variables, we use a unit system as $G=M=c=1$. Hence, the units of the radial coordinate, velocity, angular momentum, and effective potential are measured in units of $r_g$, $c$, $r_g\,c$, and $c^2$, respectively. We recall that $r_g(=\frac{G\,M}{c^2})$ is the gravitational radius.  

\begin{figure}
 \centering
 \includegraphics[width=250pt]{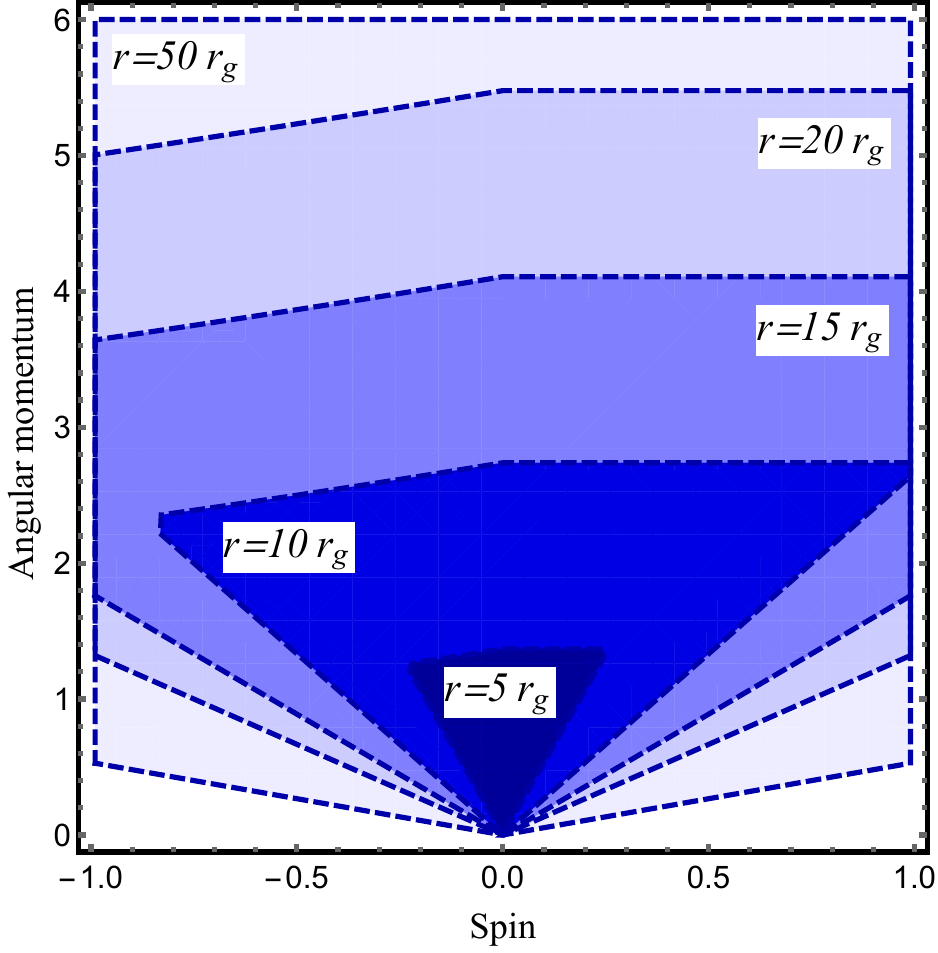}
    \caption{
The $s-\ell$ parameter space in the post-Newtonian framework. Each region shows the allowed values of the spin parameter and the dimensionless angular momentum at a specific radius. For the values beyond the specified boundaries, the post-Newtonian approximation is not reliable for that specific radius, while it still works at outer radii. For instance, for $\ell\gtrsim 4$, one can study the accretion disk at $r\gtrsim 15\, r_g$ in the post-Newtonian gravity. }
    \label{fig1}
\end{figure}

Using the above choice of the unit system, we redefine the post-Newtonian effective potential \eqref{phi-PN} as follows,
\begin{align}\label{phi-PN_dim}
\Phi_{\rm eff}^{\rm PN}=\frac{\ell^2}{2\,r^2}-\frac{1}{r}\Big[1+\frac{4}{3}\big(\frac{\ell^2}{r^2}-\frac{s \ell}{r^2}\big)\Big].
\end{align}
This relation indicates that the relativistic part of the effective potential is a function of $r^{-3}$, while the standard one is a function of $r^{-1}$ and $r^{-2}$. Then, as the radius decreases, this correction grows faster. However, at each radius, the value of this correction should not exceed the Newtonian term, and the ratio of corrections to the standard term should be very small. We assume this ratio is at most of the order $10^{-1}$. The allowed $s-\ell$ parameter space in which the above condition is satisfied is exhibited in Fig. \ref{fig1}. As shown, at the inner radii, the allowed region reasonably shrinks. Here, we consider the interval $-1\leq s \leq 1$ for the spin parameter \cite{2016NewA...43...10S}. The lower (upper) limit corresponds to the maximally rotating retrograde (prograde) black hole with an accretion disk.
It should be mentioned that considering some interactions of the accretion disk with the central body, in some works, for instance in \cite{1974ApJ...191..507T}, it is shown that the upper limit of the spin parameter is $0.998$. In the present work, we do not consider such effects and set the upper limit to $1$.

\begin{figure}
    \centering
    \includegraphics[width=400pt]{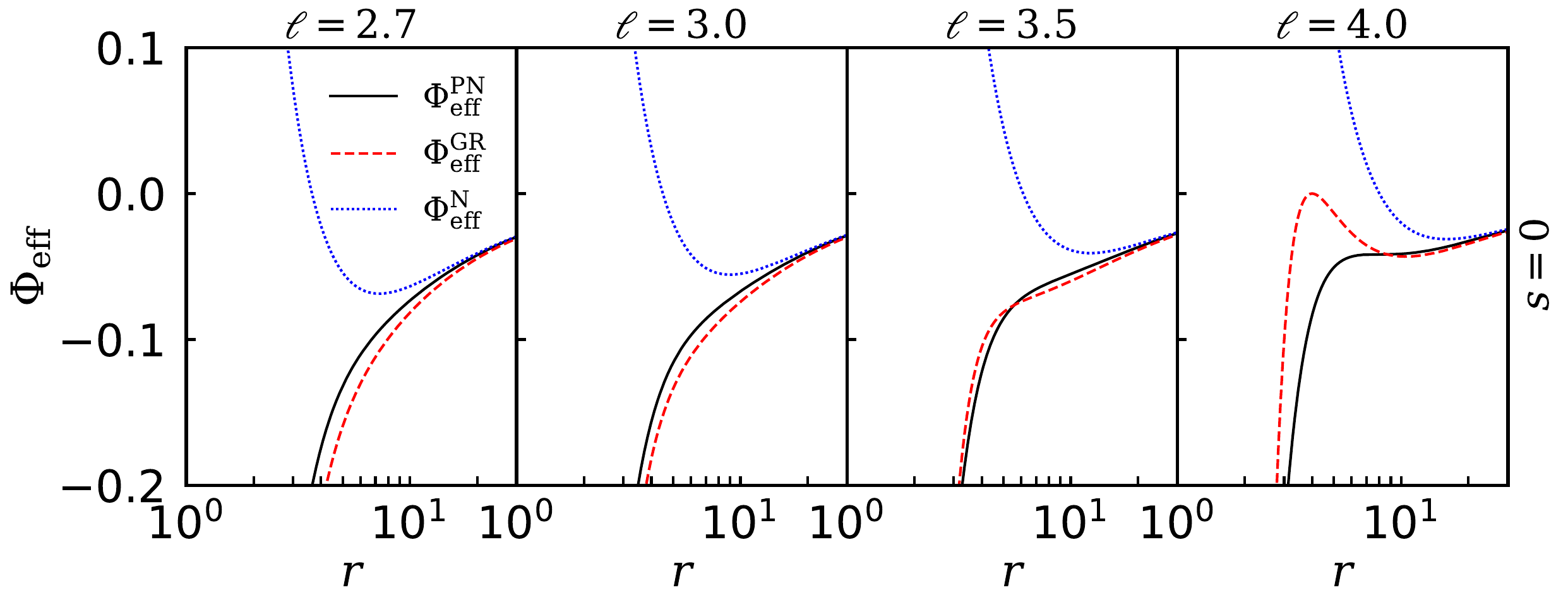} 
    \includegraphics[width=400pt]{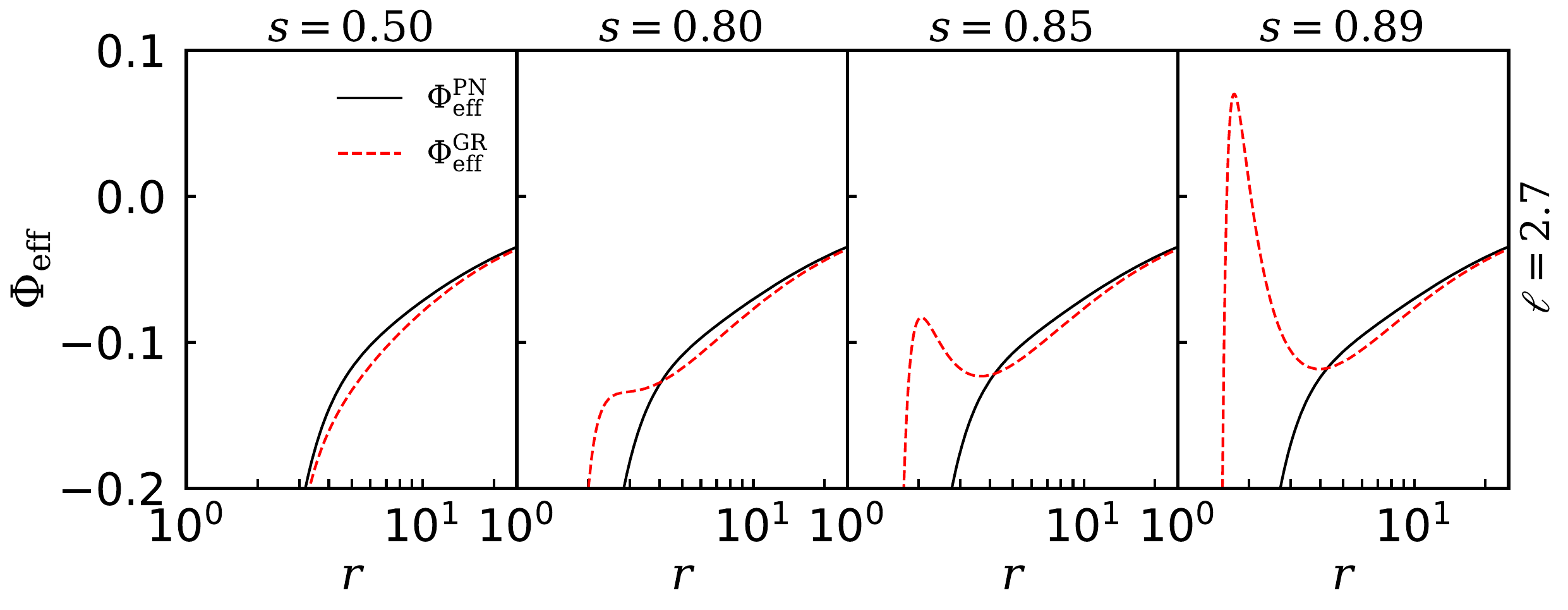}
    \includegraphics[width=400pt]{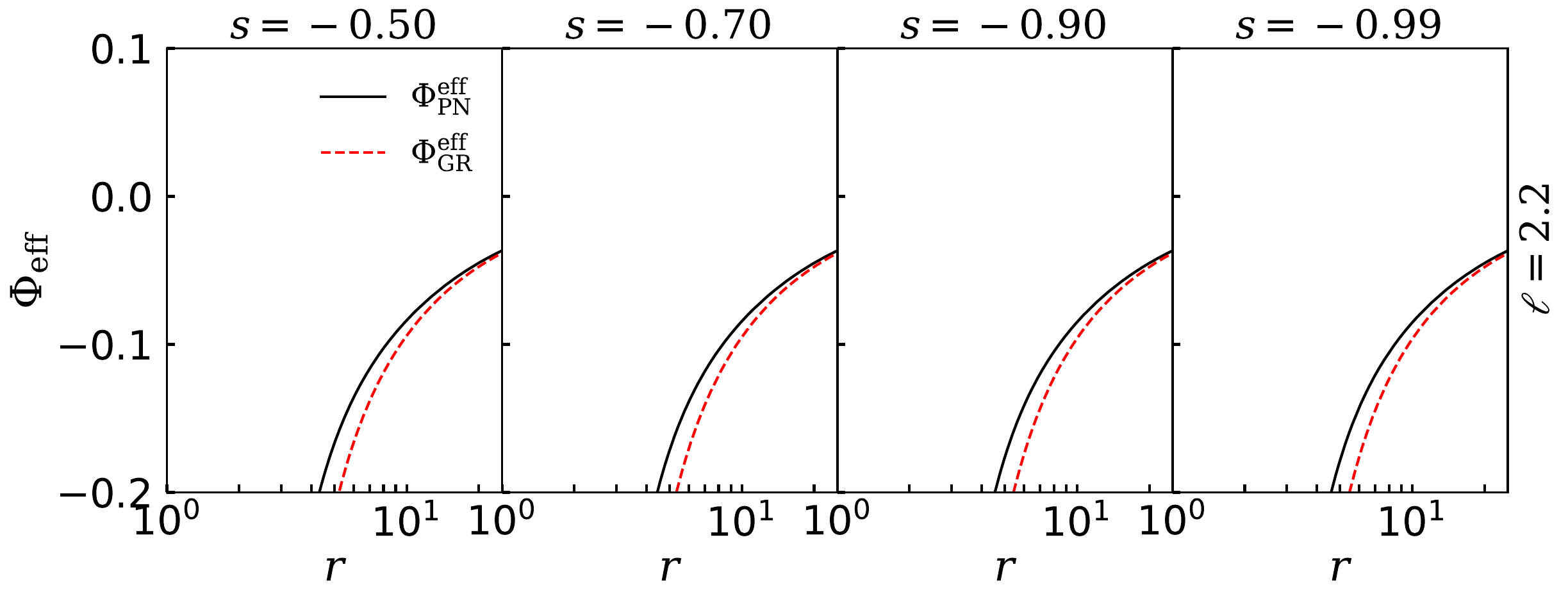}
    \caption{Comparison of effective potentials ($\Phi_{\rm eff}$): Post-Newtonian ($\Phi_{\rm eff}^{\rm PN}$), general-relativistic ($\Phi_{\rm eff}^{\rm GR}$) and Newtonian ($\Phi_{\rm eff}^{\rm N}$), as a function of logarithmic radial distance ($r$) for different $\ell$ and $s$ is presented here. In the upper panel, we fix spin parameter $s=0$ and vary the constant angular momentum as $\ell=2.7, 3.0, 3.5, 4.0$, respectively. Here, solid, dashed, and dotted curves correspond to post-Newtonian, GR, and Newtonian cases. In the middle panel, $\Phi_{\rm eff}^{\rm PN}$ and $\Phi_{\rm eff}^{\rm GR}$ are plotted for a fixed $\ell=2.7$, and the spin parameter is only varied here $s = 0.50,0.80,0.85$, and $0.89$. Finally, in the lower panel, we repeat the above exercise for the retrograde spin as the middle panel, only the spin orientation is taken as opposite as, $s=-0.50, -0.70, -0.90, -0.99$ and $\ell=2.2$.}
    \label{fig:pot}
\end{figure}

Given this parameter space, we examine the behavior of $\Phi_{\rm eff}^{\rm PN} (\equiv \Phi_{\rm SR} = \Phi_{\rm SN})$ and compare it with the exact GR and the Newtonian cases (see Fig. \ref{fig:pot}). The Newtonian effective potential is
\begin{align}\label{Phi_eff_N}
\Phi_{\rm eff}^{\rm N}=\frac{\ell^2}{2\,r^2}-\frac{1}{r},
\end{align}
as well as the GR one introduced by \cite{2018PhRvD..98h3004D} is given as
\begin{align}
\Phi_{\rm eff}^{\rm GR}= \frac{1}{2} \ln \Bigg[\frac{r (r^2 + s^2 - 2r)}{s^2 (r+2)-4s\ell + r^3 - \ell^2 (r-2)}\Bigg].
\end{align}

In the upper panels of Fig. \ref{fig:pot}, the value of $\ell$ is chosen from the region of Fig. \ref{fig1}, which is valid for $r> 10\,r_g$ and $s=0$.  
These panels show where the relativistic effects turn on, the Newtonian case starts to deviate from GR, although the post-Newtonian potential remains to mimic the relativistic systems. Reasonably, $\Phi_{\rm eff}^{\rm PN}$ is sandwiched between $\Phi_{\rm eff}^{\rm N}$ and $\Phi_{\rm eff}^{\rm GR}$. 
Interestingly, the post-Newtonian results work well even outside the realm of its validity, $5\,r_g<r<10\,r_g$.  This behavior of the post-Newtonian result observed here is reminiscent of the ``unreasonable'' effectiveness of post-Newtonian approximation reported in gravitational wave physics, cf. \cite{2011PNAS..108.5938W}. 
In the middle row, we choose $s>0$ and pick up a suitable $\ell=2.7$ from Fig. \ref{fig1}. As seen, the deviation between the post-Newtonian and GR cases appears at highly relativistic radii. This deviation grows by increasing the spin parameter. However, the bottom row panels ($s<0$) reveal that there is no such difference between $\Phi_{\rm eff}^{\rm PN}$ and $\Phi_{\rm eff}^{\rm GR}$ even at $r<5\,r_g$. In fact, in these cases, the post-Newtonian results \textit{unreasonably} mimic the GR results.  

\subsection{Critical point analysis} \label{transonic-method}

Accretion flow around compact objects is generally transonic in nature, where the accreting matter gets trapped under the immense gravitational potential. In reality, the convergent flow enters far from the central object ($r\sim 1000\,r_g$), either from the binary companion or the surrounding media, with a negligible radial velocity ($V\ll c$; subsonic). However, the flow changes its sonic state by overcoming the local sound speed, \textit{\textit{i.e.,}} $V>C_{\rm s}$ at a certain radius, usually known as the critical point ($r_{\rm c}$). At $r=r_{\rm c}$, the wind equation \eqref{wind-SR}, takes an indeterminate form \textit{\textit{i.e.,}} $\frac{dV}{dr}|_{r_{\rm c}}= \frac{0}{0}$, where the conditions $N(r_{\rm c},V_{\rm c},C_{{\rm s}_{\rm c}},\ell_{\rm c},s)=0$ and $D(r_{\rm c},V_{\rm c},C_{{\rm s}_{\rm c}},\ell_{\rm c},s)=0$ are known as the critical point conditions. Therefore, we need to use the l'H$\hat{{\rm o}}$pital rule to avoid such discontinuity and make the velocity gradient smooth. 
\begin{align*}
    &\frac{dV}{dr}|_{\rm c} = \frac{{dN}/{dr}|_{\rm c}}{ {dD}/{dr}|_{\rm c}}\\ &= \frac{(\frac{\partial N}{\partial r})_{\rm c} + (\frac{\partial N}{\partial V})_{\rm c} \, (\frac{dV}{dr})_{\rm c}+ (\frac{\partial N}{\partial C_{\rm s}})_{\rm c} \, (\frac{dC_{\rm s}}{dr})_{\rm c} + (\frac{\partial N}{\partial \ell})_{\rm c} \, (\frac{d\ell}{dr})_{\rm c}}{(\frac{\partial D}{\partial r})_{\rm c} + (\frac{\partial D}{\partial V})_{\rm c} \, (\frac{dV}{dr})_{\rm c}+ (\frac{\partial D}{\partial C_{\rm s}})_{\rm c} \, (\frac{dC_{\rm s}}{dr})_{\rm c} + (\frac{\partial D}{\partial \ell})_{\rm c} \, (\frac{d\ell}{dr})_{\rm c}}.
\end{align*}
We replace the radial derivatives of angular momentum and sound speed with the corresponding gradient equations, \eqref{ell-SR}, \eqref{temp-grad-SR}, and eventually obtain a quadratic equation for $dV/dr|_{\rm c}$. Hence, depending on the flow parameters, $dV/dr |_{\rm c}$ can take two different values, and the nature of the critical point is determined. It is noteworthy that the presence of flow angular momentum ($\ell$) introduces the notion of multiple critical points (MCPs). The nomenclature of such critical points is estimated according to the distance from the compact object: inner ($r_{\rm c}^{\rm in}$; near to the compact object), middle ($r_{\rm c}^{\rm mid}$) and outer ($r_{\rm c}^{\rm out}$; far from the central object), respectively. Among these, $r_{\rm c}^{\rm in}$, $r_{\rm c}^{\rm out}$ are the physical (`X' or saddle type: $\frac{dV}{dr}|_{\rm c}$ is real) ones through which a transonic flow can only pass thorough. However, we are interested in the outer critical point passing solutions, as the present post-Newtonian regime is valid outside $10\,r_{g}$ (see Sec. \ref{sec2} for details). Further, we find two real roots for the `X'-type critical points: $\frac{dV}{dr}|_{r_{\rm c}}<0$, which corresponds to accretion, and the other $\frac{dV}{dr}|_{r_{\rm c}}>0$ refers to the wind solutions (see \cite{2022MNRAS.516.5092M,2023MNRAS.523.4431M} and references therein). It is also important to note that the flow possessing MCPs can potentially harbor shock waves \cite{2018MNRAS.473.2415S,2018PhRvD..98h3004D,2022MNRAS.514.1940D}, such scenarios are beyond the scope of this paper.
\subsection{Global accretion solutions}
In order to determine the global transonic solution for the steady, thin, axisymmetric, low-$\ell$ accretion flow in the post-Newtonian space-time near a spinning compact object, one must solve a set of coupled differential equations. These include the wind equation, the azimuthal momentum equation, and the temperature gradient equation (as detailed in \cite{2007MNRAS.376.1659D,2022MNRAS.516.5092M,2023MNRAS.523.4431M}). 
We aim to focus only on the outer critical point ($r_{\rm c}^{\rm out}$) passing solutions as the 1\tiny PN \normalsize correction works smoothly outside $10\,r_g$.

\begin{figure}
    \centering
    \includegraphics[width=400pt]{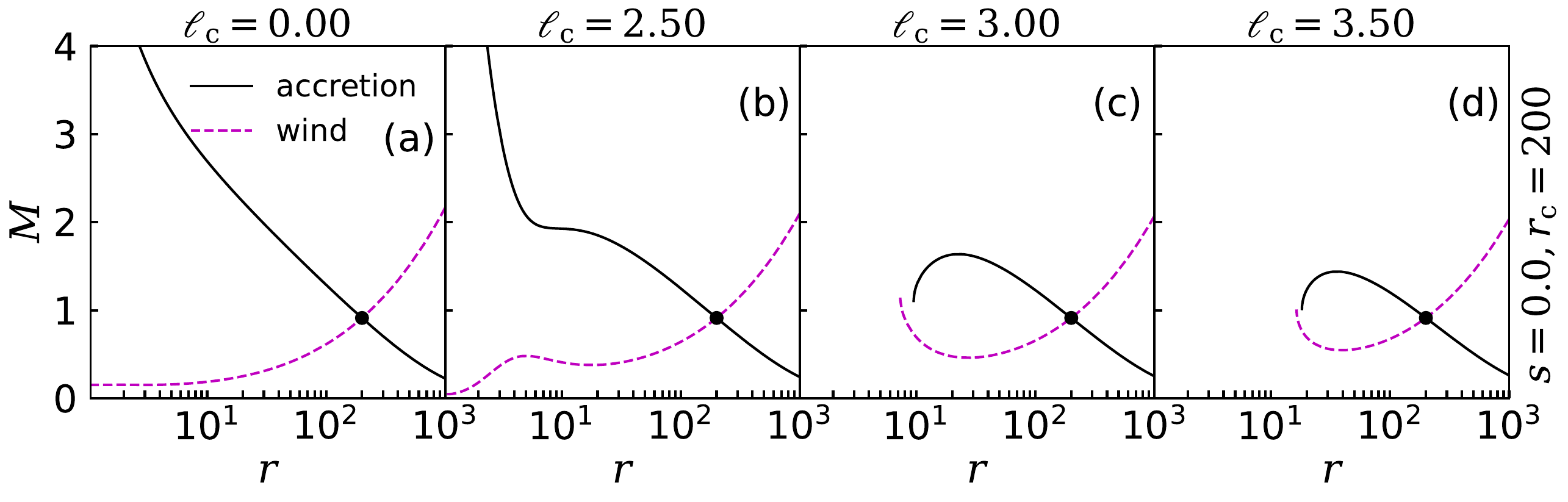} 
    \quad \includegraphics[width=400pt]{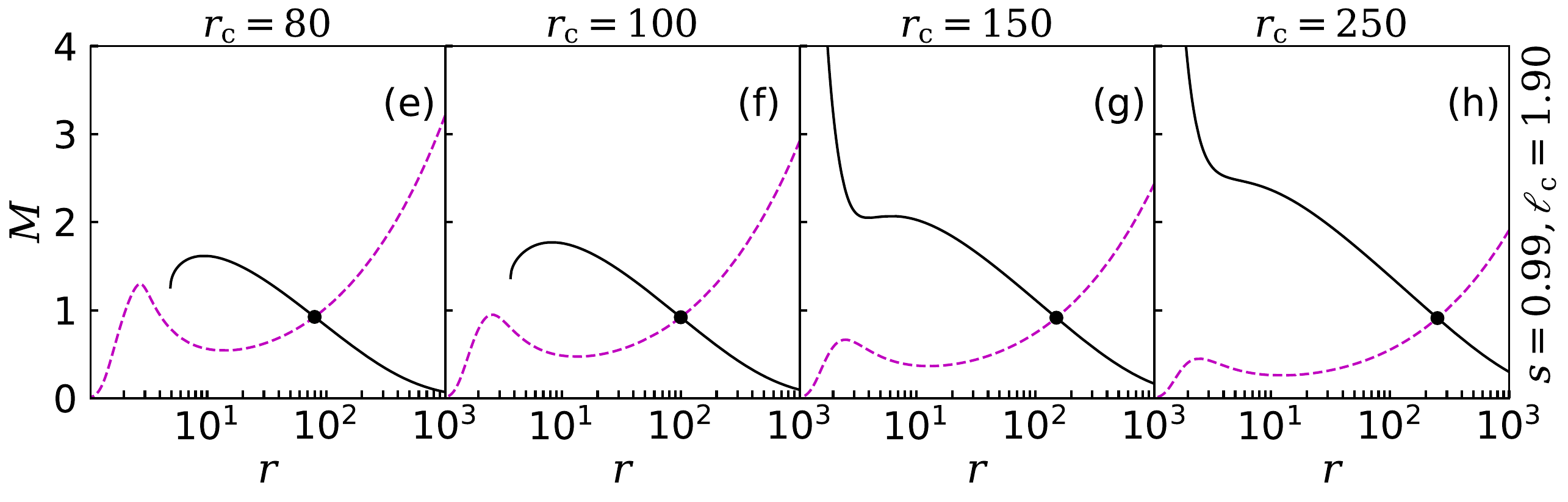} 
    \quad \includegraphics[width=400pt]{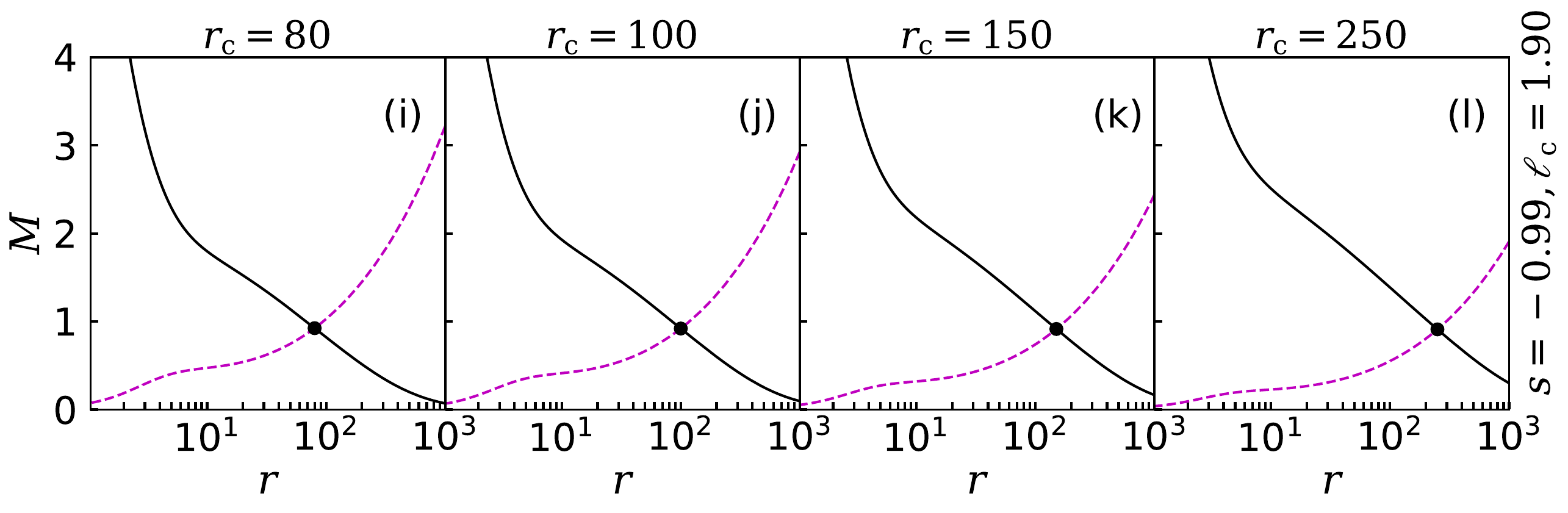} 
    \caption{Upper panel: Mach number ($M=V/C_{\rm s}$) for the semi-relativistic (SR) fluid is plotted as a function of radial distance for $s=0$ in the post-Newtonian framework. Here, we fix the outer critical point ($r_{\rm c}$) at 200, and the angular momentum at the critical point ($\ell_{\rm c}$) is only varied as $\ell_{\rm c}=0$ (Bondi-type), 2.50, 3.00, and 3.50, respectively. Lower panel: Here, we fix the spin of the compact object as, $s=0.99$ and vary the critical point location as $r_{\rm c}=80, 100, 150$, and 250. Note the solid curve corresponds to accretion solutions, whereas the dashed curves represent wind solutions, and the filled circle corresponds to the critical point location.}
    \label{fig:Mach_SR}
\end{figure}

In Fig. \ref{fig:Mach_SR}, we present a family of global accretion solutions, where Mach number ($M=V/C_{\rm s}$) is plotted as a function of logarithmic radial distance ($r$) for various input parameters, ($r_{\rm c},s,\ell_c$). To begin with, we choose the input flow parameters as ($r_{\rm c},s,\ell_{\rm c}$) = ($200,0,0$) and eventually solve the critical point conditions $N_{\rm SR}=D_{\rm SR}=0$ at $r_{\rm c}$ to obtain $dV/dr|_{\rm c}$ (see Fig. \ref{fig:Mach_SR}a). Once we find the velocity gradient, we integrate the flow variables ($V, C_{\rm s}, \ell$) by solving Eqs. \eqref{ell-SR}, \eqref{wind-SR}, and \eqref{temp-grad-SR} both towards and away from the compact object (\textit{i.e.,} upto the outer edge of the disk, $r_{\rm edge}=1000\,r_g$). For the given set of input parameters, the flow smoothly connects the post-Newtonian radius $r_{\rm PN}=10 \, r_g$, to the $r_{\rm edge}$ for the SR regime. This particular case where $\ell=0$ is known as the `Bondi-type' solution. A recent study \cite{2021PhRvD.104b4056K} also showcased the spherically symmetric accretion in the post-Newtonian limit. We further introduce the rotation ($s$) of the central object as well as the flow angular momentum ($\ell$) in our analysis, which makes this study more general.

Remember, as pointed out in Fig. \ref{fig:pot}, the effective potential mimics the GR potential even beyond the post-Newtonian radius ($r<r_{\rm PN}$), encouraging us to explore the solutions even in this region. Interestingly, the Mach number monotonically increases in this regime, which is similar to the GR case, see \cite{2022MNRAS.516.5092M}. Next, we increase $\ell_{\rm c}=2.50$ and find an open-type global accretion solution like the previous one. Further, when we increase $\ell_{\rm c}$ to $3.00$, the inflowing matter entering from the outer edge just touches the $r_{\rm PN}$ but becomes the closed-type solution. Further, at a higher $\ell_{\rm c}=3.50$, the solution fails to reach the post-Newtonian radius and gets a closed topology (Fig. \ref{fig:Mach_SR}c,d). Such kinds of solutions are of less physical importance as they fail to reach near to the compact object. 

In the middle panels, we plot the radial profile of Mach number ($M$) for the maximally spinning case, $s=0.99$, and keep $\ell_{\rm c}=1.90$ in all the panels of Fig. \ref{fig:Mach_SR}e-h. Here, we vary the location of the outer critical point from $r_{\rm c}=80,100,150$, up to $200$, respectively. For the first two cases, we obtain accretion solutions that are closed in nature, however, the flow reaches up to $r_{\rm PN}$. Although the accretion solutions remain closed-type, we find that the wind branch reaches smoothly up to the inner edge, and these kinds of solutions are reported earlier by \cite{2004IJMPD..13.1955D} for the case of black holes. Interestingly, as we increase $r_{\rm c}$ beyond 150, we get open-type global transonic solutions that connect the inner edge of the disk with the $r_{\rm edge}$. Finally, in the lower panels (i-l) we plot the accretion solutions for the retrograde case, \textit{i.e.,} $s=-0.99$. In all these cases, we fix $\ell_{\rm c}=1.90$, and the critical point location is varied as in the middle panels. For each of the panels in Fig. \ref{fig:Mach_SR}i-l, we get open-type global; transonic accretion solutions. 

It should be noted that the present post-Newtonian framework works reasonably well up to $r_{\tiny {\rm PN}}=10\,r_g$. Because of this, the global accretion solutions in the post-Newtonian framework always possess a single outer critical point only (see Fig. 3). However, this is not the case for the GR-framework, as we can have both inner and outer critical points passing global accretion solutions \cite[and references therein]{2018PhRvD..98h3004D,2022MNRAS.516.5092M}. Hence, the  post-Newtonian framework fails to harbor shock waves, which is commonly observed in the GR context.

\begin{figure}
    \centering
    \includegraphics[width=300pt]{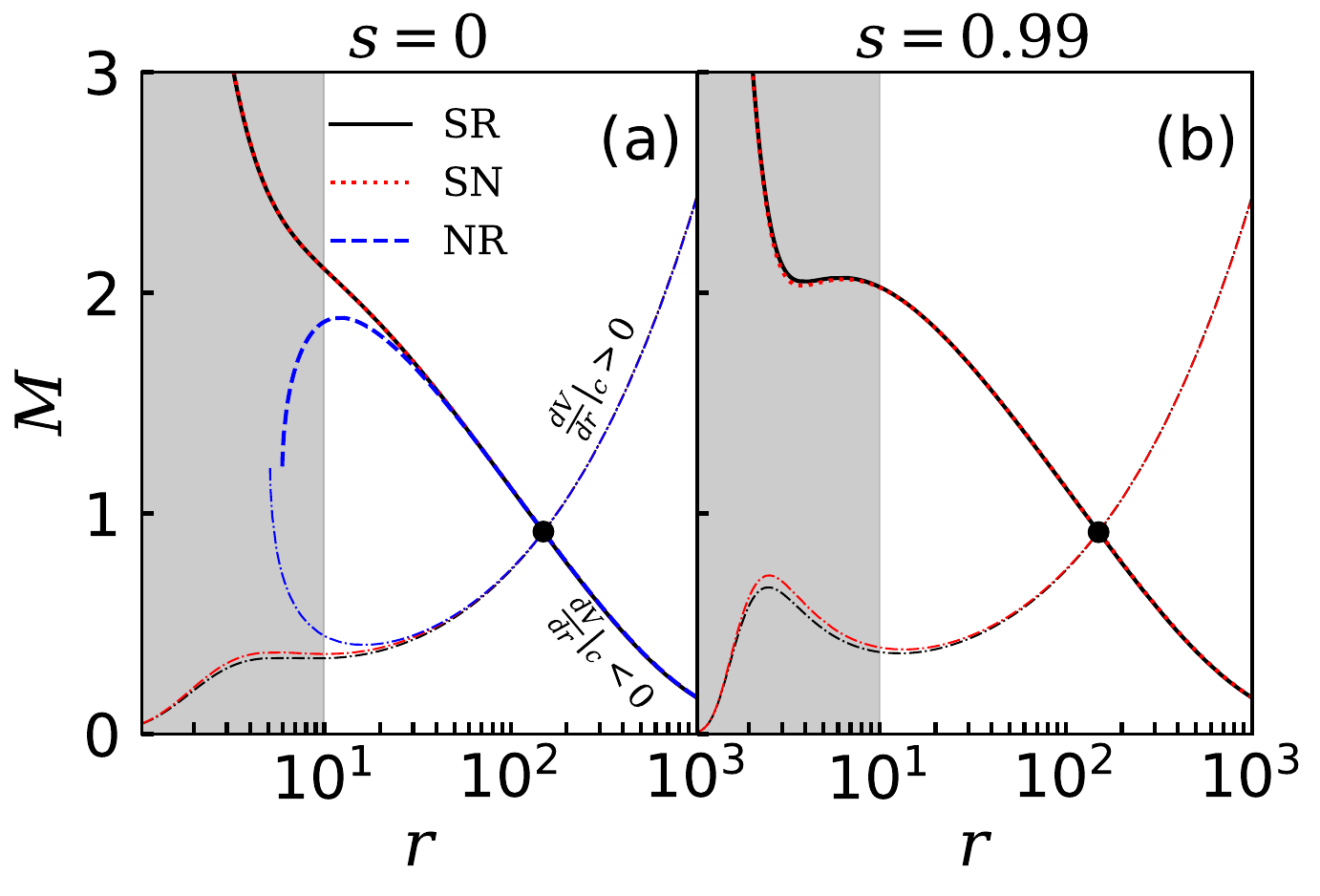}
    \caption{Comparison of accretion for SR (solid), SN (dotted), and NR (dashed) cases, respectively. We fix the spin $s=0$ in the left panel, whereas the right panel corresponds to $s=0.99$. In both panels, $r_{\rm c}=200$ and $\ell_{\rm c}=1.90$ are kept fixed. Here, $\frac{dV}{dr}|_{\rm c}<0$ branch refers to the accretion solution (solid curves), whereas $\frac{dV}{dr}|_{\rm c}>0$ (dot-dashed curves) represents the wind branch. The gray-shaded region represents $r<r_{\rm PN}$.}
    \label{fig:SN_SR_NR}
\end{figure}

In the next, we compare the accretion solutions among the SR, SN, and NR cases in Fig. \ref{fig:SN_SR_NR}. Note that, for the NR approximation, $\ell$ is a constant and terms containing spin ($s$) vanishes. We therefore choose $(r_{\rm c},s,\ell_{\rm c})=(200,0,1.90)$ in the left panel. We find that the SR and SN results match closely all over the radius, but the NR one deviates from the other two around the radius, $r\sim 30$. In the right panel, we include the maximally spinning case $s=0.99$, keeping other parameters fixed. Here, we notice similar behavior in SR and SN results. Interestingly, we find the low angular momentum accretion solutions in the NR limit fail to provide an open-type solution for any choice of input parameters. This fact also illustrates the importance of the relativistic/post-Newtonian corrections.

\begin{figure}
    \centering
    \includegraphics[width=300pt]{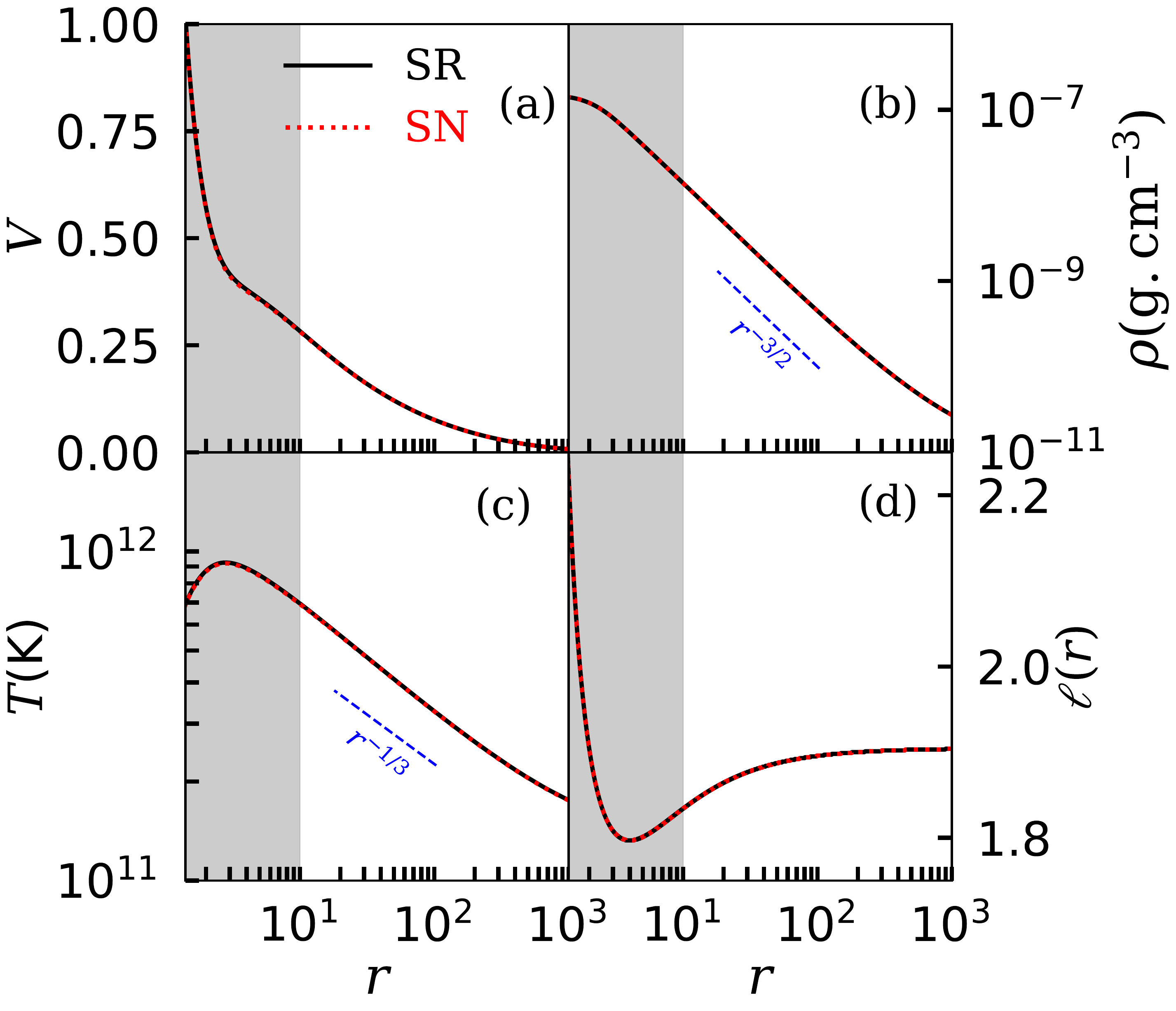}
    \caption{A comparative study of the flow variables, namely the velocity ($V$), mass-density ($\rho$), temperature ($T$), and the angular momentum ($\ell$) for the set of solutions obtained in Fig. \ref{fig:SN_SR_NR}b are plotted as a function of radial coordinate ($r$). Here, the gray-shaded region ($r<r_{\rm PN}$) is beyond the scope of the post-Newtonian framework. Note that we depict only the accretion solutions, not the wind, as per the point of interest.}
    \label{fig:flow_varibles}
\end{figure}

Further, in Fig. \ref{fig:flow_varibles}, we explore the variation of the flow variables \textit{i.e.,} (a) velocity, $V$, (b) the mass-density, $\rho$, (c) temperature, $T=\frac{2 m_{\rm p} C_{\rm s}^2 }{k_{\rm B}}c^2$, where $m_{\rm p}$ is the proton mass and $k_{\rm B}$ is the Boltzmann's constant, and (d) flow angular momentum, $\ell$, for the SR and SN cases with $s=0.99$ and $\ell_{\rm c}=1.90$. Here, $r_{\rm c}$ is kept at 200 (same as in the right panel of Fig. \ref{fig:SN_SR_NR}). We observe that the sub-sonic accretion flow from the outer edge of the disk ($r_{\rm edge}=1000$) eventually gains its radial velocity as it moves inwards and makes a smooth transition to become super-sonic at the outer critical point ($r_{\rm c}=200$) before falling into the central body. 
We depict the density profile of the converging flow in Fig. \ref{fig:flow_varibles}b, where a gradual increase of density is observed with decreasing $r$. This happens mainly due to the geometric compression of infalling matter, and as a consequence, the temperature of the flow is expected to increase with the decrease of radial distance, as shown in panel (c).

We notice the disk temperature reaches $T \gtrsim 6.5\times 10^{11}$ K even within the post-Newtonian regime, $r>r_{\rm PN}\equiv 10\,r_g$. It is noteworthy that the density profile follows the exact same power-law distribution $\rho \propto r^{-3/2}$ as in the self-similar limit, whereas the temperature profile follows a shallower one $T\propto r^{-1/3}$. However, for the global accretion solutions in GR, flow temperature varies as, $T \propto r^{-3/4}$. Further, in panel (d), we depict the radial profile of flow angular momentum, and it shows a slowly decreasing behavior up to $r_{\rm PN}$ (see Appx. \ref{discussion} for details). This can be due to the presence of the spin-orbit coupling in the governing post-Newtonian equations. This feature was not reported earlier in the post-Newtonian accretion.

\section{Concluding remarks}\label{sec6}

In this paper, we have shown the behavior of global transonic accretion solutions in the post-Newtonian framework up to the first post-Newtonian (1\tiny PN \normalsize) order. We have formulated a detailed analysis of the post-Newtonian hydrodynamics around a spinning compact body. We have obtained the explicit relativistic hydrodynamic equations that allow one to study accretion disks around rotating objects without imposing any symmetries up to the 1\tiny PN \normalsize order. 
The aim is to bridge the gap between the Newtonian framework to exact GR solutions. In doing so, we have considered the disk to be confined on the equatorial plane (\textit{i.e.,} $\theta = \pi/2$), and axisymmetry has been taken into account.
The equations \eqref{Euler_r}, \eqref{Euler_phi}, and \eqref{Euler_z} describe the fluid motion around a spinning compact body in the post-Newtonian gravity where several relativistic corrections to the Newtonian equations, gather together in the braces, play a role. Among them, an interesting relativistic correction to the hydrodynamic equations has been obtained here\textemdash the spin-orbit interaction. This effect is not present in Newtonian gravity. It should be noted that removing these corrections reproduces the predictions given in Newtonian gravity. 
Further, we have explored all possibilities for the small limits in radial ($V$) and azimuthal velocities ($\beta_\phi$), thermal energy ($\Pi$) by analyzing three case studies: Semi-relativistic (SR), Semi-Newtonian (SN) and Non-relativistic (NR) in the steady-state limit ($\partial/\partial{t} \rightarrow 0$). Our findings are as follows:

\begin{itemize}
    \item We compute the allowed range of flow angular momentum ($\ell$) for the allowed range of spin ($s$) values (see Fig. \ref{fig1}), which bounds us to choose the appropriate $\ell$ from the region of interest, \textit{i.e.,} $r>r_{\rm PN}$.  However, we are interested in low-angular momentum flows only. Further, with the suitable $\ell$, we compare the effective potentials among the post-Newtonian, GR, and Newtonian cases. We notice from Fig. \ref{fig:pot} that the post-Newtonian potential tends to follow the exact GR from infinity up to $r_{\rm PN}$ while the Newtonian potential starts to deviate dramatically from the relativistic cases around $r_{\rm PN}$. This indicates that compared to the Newtonian analysis, the relativistic corrections considered in the post-Newtonian gravity provide us with a more realistic description. This figure also reveals that in some cases, the post-Newtonian results \textit{unreasonably} mimic the GR results in highly relativistic regions. However, this behavior is not reliable because it is beyond the realm of validity of the post-Newtonian gravity. To investigate a system at inner radii, \textit{i.e.,} $r<r_{\text{PN}}$, one needs to take into account the exact relativistic corrections and apply the Kerr metric. This is indeed one of the main shortcomings of the post-Newtonian approximation. Furthermore, in the post-Newtonian framework, to improve the measurements in the post-Newtonian zone, \textit{i.e.,} $r_{\text{PN}} \leqslant r \leqslant r_{\infty}$, one should consider higher corrections and utilize the post-Newtonian expansion of the metric containing at least the 2\tiny PN \normalsize corrections which is complicated task.  

    \item We obtain a complete set of global transonic accretion solutions for the first time for a rotating compact object in the post-Newtonian framework (see Fig. \ref{fig:Mach_SR}). Excellent agreement is noticed from both SR and SN limits (see Fig. \ref{fig:SN_SR_NR}). However, the NR (Newtonian) limit fails to provide an open-type solution. This fact illustrates the importance of the post-Newtonian corrections. 
    It should be noticed that as indicated in Fig. \ref{fig:SN_SR_NR}, these global transonic accretion solutions obtained in the post-Newtonian gravity are not reliable at the inner radii, and the exact GR should be applied. In fact, the post-Newtonian framework fails to harbor shock waves, which is commonly observed in the GR context \cite{2012NewA...17..254D}.

    \item We observe that the density profile follows the self-similar ADAF limit, \textit{i.e.,} $\rho \propto r^{-3/2}$ \cite{1994ApJ...428L..13N}. However, the temperature profile is shallower ($T\propto r^{-1/3}$) than the general relativistic (GR) profiles $T \propto r^{-3/4}$.  
    Fig. \ref{fig:flow_varibles}a exhibits the radial velocity $V$ of a fluid element in terms of radius. As seen, it grows with decreasing $r$ and reaches relativistic values close to the compact body. This is where the post-Newtonian approximation fails.
    So, the current analysis cannot be applied to investigate the velocity of the particles/fluid elements reaching the surface of the compact body. It is one of the limitations of the present study. 
     We also notice the radial variation in the angular momentum profile (see Fig. \ref{fig:flow_varibles}d) even in the absence of any viscosity/magnetic fields. 
    In Appx. \ref{discussion}, we show that even if there is no dissipative effect in the system, the pure gravitational effects can remove angular momentum from the system. 
\end{itemize}

With all the above findings, we wish to emphasize that the post-Newtonian hydrodynamics successfully mimic the exact solutions up to $r\sim r_{\rm PN}$. Strictly speaking, the current analyses, restricted to the 1\tiny PN \normalsize order, are reliable from very far radii, $r_{\infty}$, up to $r_{\text{PN}}$. However, we ignore the self-gravity for simplicity, which might be useful in the context of gravitational waves from the disk. As an introductory approach, we adopt the SR, SN, and NR limits for the purpose of simplicity, although accretion solutions involving full post-Newtonian corrections are more suitable, which we plan to consider for future endeavors.
Furthermore, to determine whether the post-Newtonian approximation can provide reasonable results from an observational point of view, we plan to study the quasi-periodic oscillations (QPOs) in the post-Newtonian accretion disk.

\section*{Acknowledgments}
The Authors thank Mahmood Roshan for reading the manuscript and providing constructive comments. 
EN gratefully acknowledges the support of Ferdowsi University of Mashhad. SM acknowledges the prestigious Prime Minister's Research Fellowship (PMRF), Government of India, for the financial support. SM is indebted to Prof. Patrick Das Gupta for an illuminating discussion on post-Newtonian theory. SD thanks Science and Engineering Research Board (SERB), India for support under grant MTR/2020/000331. 
Worthwhile comments by the anonymous referee are also gratefully acknowledged.

\bibliographystyle{apsrev4-1}
\bibliography{short,Accretion_PN_1}

\appendix

\section{Contravariant components of metric and Christoffel symbols}\label{app1}

This appendix is dedicated to introducing the essential quantities for performing the post-Newtonian calculations.   
The first are the contravariant components of the metric:
\begin{subequations}
\begin{align}
&g^{00}=-1-\frac{2}{c^2}\frac{G\,M}{r}-\frac{2}{c^4}\Big(\frac{G\,M}{r}\Big)^2+O(c^{-6}),\\
&g^{0j}=\frac{2}{c^3}\frac{G(\bm{x}\times\bm{S})^j}{r^3}+O(c^{-5}),\\
&g^{jk}=\Big(1-\frac{2}{c^2}\frac{G\,M}{r}\Big)\delta^{jk}+O(c^{-4}).
\end{align}
\end{subequations}
In order to simplify the energy-momentum conservation \eqref{con_EM} and study its zeroth and spatial components \eqref{eq_E} and \eqref{s_com}, we need to obtain the Christoffel symbols $\Gamma^{\mu}_{\alpha\beta}$ from Eq. \eqref{g_00}-\eqref{g_jk}. Using the relation $\Gamma^{\mu}_{\alpha\beta}=\frac{1}{2}g^{\mu\nu}\Big(\partial_\alpha g_{\nu\beta}+\partial_\beta g_{\nu\alpha}-\partial_\nu g_{\alpha\beta}\Big)$, after some manipulations, and truncating the results to the required post-Newtonian order, we arrive at
\begin{subequations}
\begin{align}
&\Gamma^{0}_{00}=\frac{1}{c^3}\frac{G\,M}{r^2}\partial_{t}r+O(c^{-5}),\\
&\Gamma^{0}_{0j}=\frac{1}{c^2}\frac{G\,M}{r^2}\partial_{j}r+O(c^{-4}),\\\nonumber
&\Gamma^{0}_{jk}=-\frac{1}{c^3}\bigg\lbrace\frac{G\,M}{r^2}\partial_{t}r\delta_{jk}-\frac{G}{r^4}\Big[3\Big(\big(\bm{x}\times\bm{S}\big)_j\partial_k r+\big(\bm{x}\times\bm{S}\big)_k\partial_jr\Big)\\
&-r\Big(\big(\partial_j\bm{x}\times\bm{S}\big)_k+\big(\partial_k\bm{x}\times\bm{S}\big)_j\Big)\Big]\bigg\rbrace+O(c^{-5}),\\\nonumber
&\Gamma^{j}_{00}=\frac{1}{c^2}\frac{G\,M}{r^2}\partial_j r+\frac{2}{c^4}\frac{G}{r^4}\bigg\lbrace r\Big[\big(\bm{x}\times\partial_t \bm{S}\big)_j+\big(\partial_t\bm{x}\times \bm{S}\big)_j\Big]\\
&-3\partial_t r\big(\bm{x}\times\bm{S}\big)_j\bigg\rbrace-\frac{4}{c^4}\frac{G^2M^2}{r^3}\partial_jr+O(c^{-6}),\\\nonumber
&\Gamma^{j}_{0k}=-\frac{1}{c^3}\bigg\lbrace\frac{G\,M}{r^2}\partial_tr\delta_{jk}-\frac{G}{r^4}\Big[3\Big(\big(\bm{x}\times\bm{S}\big)_k\partial_j r-\big(\bm{x}\times\bm{S}\big)_j\partial_k r\Big)\\
&-r\Big(\big(\partial_j\bm{x}\times\bm{S}\big)_k-\big(\partial_k\bm{x}\times\bm{S}\big)_j\Big)\Big]\bigg\rbrace+O(c^{-5}),\\
&\Gamma^{j}_{kn}=-\frac{1}{c^2}\frac{G\,M}{r^2}\bigg\lbrace\delta_{jn}\partial_k r+\delta_{jk}\partial_n r-\delta_{kn}\partial_j r\bigg\rbrace+O(c^{-4}).
\end{align}
\end{subequations}

In the following, we attempt to obtain each term of \eqref{eq_E} separately. In this case, we examine each term to $O(c^{-1})$. This order is sufficient to provide local conservation of energy within the fluid. In this way, it would be clear why the above terms are shortened to these post-Newtonian orders. As mentioned, we assume a perfect fluid system whose energy-momentum tensor is given by Eq. \eqref{T_fluid}. Regarding the definition \eqref{rho*} and relation \eqref{gamma}, we have   
\begin{align}\label{eq1}
\frac{1}{c}\partial_t\big(\sqrt{-g}T^{00}\big)=&c\,\partial_t\rho^*+\frac{1}{c}\,\partial_t\Big[\rho^*\big(\frac{G\,M}{r}+\frac{1}{2}v^2+\Pi\big)\Big]+O(c^{-3}),
\end{align}
for the first term of Eq. \eqref{eq_E}. 
Here, $\Pi=\epsilon/\rho^*$ is the fluid's internal energy per unit mass. In a similar manner,
the rest of the terms are obtained as follows: 
\begin{subequations}
\begin{align}
\label{eq2}
&\partial_j\big(\sqrt{-g}T^{0j}\big) =c\,\partial_j\big(\rho^*v^j\big)+\frac{1}{c}\,\partial_j\Big[\rho^*v^j\big(\frac{G\,M}{r}+\frac{1}{2}v^2+\Pi\big)\Big]+\frac{1}{c}\,\partial_j\big(p\,v^j\big)+O(c^{-3}),\\
\label{eq3}
&\Gamma^{0}_{00}\big(\sqrt{-g}T^{00}\big)=\frac{1}{c}\frac{G\,M}{r^2}\rho^*\partial_t r+O(c^{-3}),\\
\label{eq4}
& 2\Gamma^{0}_{0j}\big(\sqrt{-g}T^{0j}\big) =\frac{2}{c}\frac{G\,M}{r^2}\,\rho^*v^j\partial_jr+O(c^{-3}),\\
\label{eq5}
&\Gamma^{0}_{jk}\big(\sqrt{-g}T^{jk}\big)=O(c^{-3}).
\end{align}
\end{subequations} 
Summing these terms gives Eq. \eqref{zeroth-com-cons}.  

To obtain the Euler equation including the leading relativistic corrections, we simplify each term of Eq. \eqref{s_com} to the order $c^{-2}$. For the first term in this equation, we get
\begin{align}\label{eq6}
&\frac{1}{c}\partial_t\big(\sqrt{-g}T^{0j}\big)=\partial_t\big(\mu\rho^*v^j\big)+O(c^{-4}),
\end{align}
where $\mu=1+\frac{1}{c^2}\big(\Pi+\frac{p}{\rho^*}+\frac{G\,M}{r}+\frac{1}{2}v^2\big)$. After some manipulations, we find the rest of the terms as follows:
\begin{subequations}
\begin{align}
\label{eq7}
&\partial_k\big(\sqrt{-g}T^{jk}\big)=\partial_k\big(\mu\rho^*v^jv^k\big)+\partial_j p+O(c^{-4}),\\
\nonumber
&\Gamma^{j}_{00}\big(\sqrt{-g}T^{00}\big) =\frac{G\,M}{r^2}\rho^*\partial_jr\bigg\lbrace\mu -\frac{1}{c^2}\Big(\frac{p}{\rho^*}+4\frac{G\,M}{r}\Big)\bigg\rbrace\\\label{eq8}
&+\frac{2}{c^2}\frac{G\,\rho^*}{r^4}\bigg\lbrace r\Big[\big(\bm{x}\times\partial_t\bm{S}\big)_j+\big(\partial_t\bm{x}\times\bm{S}\big)_j\Big]-3\partial_tr\big(\bm{x}\times\bm{S}\big)_j\bigg\rbrace+O(c^{-4}),\\\nonumber
& 2\Gamma^{j}_{0k}\big(\sqrt{-g}T^{0k}\big) =-\frac{2}{c^2}\frac{G\,M}{r^2}\rho^*v^j\,\partial_tr+\frac{2}{c^2}\frac{G}{r^4}\rho^*v^k\bigg\lbrace3\Big[\partial_jr\big(\bm{x}\times\bm{S}\big)_k-\partial_kr\big(\bm{x}\times\bm{S}\big)_j\Big]\\\label{eq9}
&-r\Big[\big(\partial_j\bm{x}\times\bm{S}\big)_k-\big(\partial_k\bm{x}\times\bm{S}\big)_j\Big]\bigg\rbrace+O(c^{-4}),\\
\label{eq10}
&\Gamma^{j}_{kn}\big(\sqrt{-g}T^{kn}\big)=-\frac{1}{c^2}\frac{G\,M}{r^2}\rho^*\bigg\lbrace\Big[2\partial_kr\,v^kv^j-\partial_jr\,v^2\Big]-\partial_jr\frac{p}{\rho^*}\bigg\rbrace+O(c^{-4}).
\end{align}
\end{subequations}

\section{Basis vectors of the local rest frame}\label{app2}

In this appendix, we obtain the orthonormal tetrad basis vectors of the LRF.   
As mentioned earlier, instead of describing the equations of motion in the global frame, we obtain them in the LRF. In fact, we describe physical quantities by projecting them on the orthonormal tetrad basis carried by an observer who is locally at rest with respect to the fluid element. To do so, we use the method introduced by \cite{1972ApJ...178..347B}. 

The standard form of the metric which is valid for any stationary, axisymmetric, asymptotically flat spacetime is given by
\begin{align}\label{stand-met}
ds^2=-e^{2\nu}c^2dt^2+e^{2\psi}\big(d\varphi-\omega\,c\,dt\big)^2+e^{2\mu_1}dr^2+e^{2\mu_2}d\theta^2.
\end{align}
By rewriting the post-Newtonian metric \eqref{g_00}-\eqref{g_jk} in the above standard form, one can deduce that  
\begin{subequations}
\begin{align}
& e^{2\nu}=1-\frac{2}{c^2}\frac{G\,M}{r}+\frac{2}{c^4}\Big(\frac{G\,M}{r}\Big)^2,\\
& e^{2\mu_1}=1+\frac{2}{c^2}\frac{G\,M}{r},\\
& e^{2\mu_2}=r^2\Big(1+\frac{2}{c^2}\frac{G\,M}{r}\Big),\\
& e^{2\psi}=r^2\sin^2\theta\Big(1+\frac{2}{c^2}\frac{G\,M}{r}\Big),
\end{align}
\end{subequations}
and
\begin{align}\label{omega}
\omega=\frac{1}{c^3}\frac{G\,M\,s}{r^3}+O(c^{-5}).
\end{align}
Here, it is assumed that the spin of the black hole is aligned with the $z$-axis. According to the general transformations between the LNRF and the standard one \eqref{stand-met} introduced in Eqs. (3.1)-(3.2) by \cite{1972ApJ...178..347B}, the Lorentz transformations between the LNRF and the CRF, and those between the CFR and the LRF, after doing straightforward calculations, we find   
\begin{subequations}
\begin{align}
\nonumber
& e_\mu^{~(t)}=\Big\lbrace\gamma_\text{tot}\Big(1-\frac{1}{c^2}\frac{G\,M}{r}+\frac{1}{c^4}\big(\frac{G\,M}{r}\big)^2\Big)+\frac{1}{c}\beta_{\varphi}\,\gamma_\text{tot}\,\omega\,r,~-\frac{1}{c}\beta_r\,\gamma_r\Big(1+\frac{1}{c^2}\frac{G\,M}{r}\Big),~0,\\
&~-\frac{1}{c}\beta_{\varphi}\gamma_\text{tot}\,r\Big(1+\frac{1}{c^2}\frac{G\,M}{r}\Big)\Big\rbrace,\\
& e_\mu^{~(r)}=\Big\lbrace -\frac{1}{c}\beta_r\,\gamma_\text{tot}\Big(1-\frac{1}{c^2}\frac{G\,M}{r}\Big),~\gamma_r\Big(1+\frac{1}{c^2}\frac{G\,M}{r}\Big),~0,~\frac{1}{c^2}\beta_r\,\beta_{\varphi}\gamma_\text{tot}\,r\Big(1+\frac{1}{c^2}\frac{G\,M}{r}\Big)\Big\rbrace,\\
& e_\mu^{~(\theta)}=\Big\lbrace 0,~0,~r\Big(1+\frac{1}{c^2}\frac{G\,M}{r}\Big),~0\Big\rbrace,\\
& e_\mu^{~(\varphi)}=\Big\lbrace -\frac{1}{c}\beta_\varphi\,\gamma_\varphi\Big(1-\frac{1}{c^2}\frac{G\,M}{r}\Big)-\gamma_\varphi\,\omega\,r,~0,~0,~\gamma_\varphi\,r\Big(1+\frac{1}{c^2}\frac{G\,M}{r}\Big)\Big\rbrace,
\end{align}
\end{subequations}
for the covariant LRF basis vectors. Here, $\gamma_{\text{tot}}=\gamma_r\gamma_\varphi$. In the above results, the case $\theta=\frac{\pi}{2}$ has been considered.
For the contravariant basis vectors, we also arrive at
\begin{subequations}
\begin{align}
\nonumber
& e^\mu_{~(t)}=\Big\lbrace \gamma_\text{tot}\Big(1+\frac{1}{c^2}\frac{G\,M}{r}-\frac{1}{c^4}\big(\frac{G\,M}{r}\big)^2\Big),~\frac{1}{c}\beta_r\gamma_r\Big(1-\frac{1}{c^2}\frac{G\,M}{r}\Big),~0,~\gamma_\text{tot}\,\omega\\
&+\frac{1}{c\,r}\beta_\varphi\,\gamma_\text{tot}\Big(1-\frac{1}{c^2}\frac{G\,M}{r}\Big)\Big\rbrace,\\\nonumber
& e^\mu_{~(r)}=\Big\lbrace \frac{1}{c}\beta_r\gamma_\text{tot}\Big(1+\frac{1}{c^2}\frac{G\,M}{r}-\frac{1}{c^4}\big(\frac{G\,M}{r}\big)^2\Big),~\gamma_r\Big(1-\frac{1}{c^2}\frac{G\,M}{r}\Big),~0,~\frac{1}{c}\beta_r\gamma_\text{tot}\,\omega\\
&+\frac{1}{c^2 r}\beta_r\beta_\varphi\gamma_\text{tot}\Big(1-\frac{1}{c^2}\frac{G\,M}{r}\Big)\Big\rbrace,\\
& e^{\mu}_{~(\theta)}=\Big\lbrace 0,~0,~\frac{1}{r}\Big(1-\frac{1}{c^2}\frac{G\,M}{r}\Big),0 \Big\rbrace,\\
& e^\mu_{~(\varphi)}=\Big\lbrace \frac{1}{c}\beta_\varphi\gamma_\varphi\Big(1+\frac{1}{c^2}\frac{G\,M}{r}\Big),~0,~0,~\frac{1}{c}\beta_\varphi\gamma_\varphi\omega+\frac{1}{r}\gamma_\varphi\Big(1-\frac{1}{c^2}\frac{G\,M}{r}\Big)\Big\rbrace.
\end{align}
\end{subequations}

Therefore, considering these vectors and the relations $u_{\mu}=e_\mu^{~(\nu)}u_{(\nu)}$, we obtain that
\begin{align}
\nonumber
&\Big(u_t,~u_r,~u_\theta,~u_\varphi\Big)=\bigg(-c\,\gamma_\text{tot}\Big(1-\frac{1}{c^2}\frac{G\,M}{r}+\frac{1}{c^4}\big(\frac{G\,M}{r}\big)^2\Big)\\
&-\ell\omega,~V\Big(1+\frac{1}{2}\frac{V^2}{c^2}+\frac{1}{c^2}\frac{G\,M}{r}\Big),~0,~\ell\bigg).
\end{align}
Here, $u_\varphi\coloneqq\ell$ is the angular momentum. Using the $\varphi$ component of $u_{\mu}=e_\mu^{~(\nu)}u_{(\nu)}$, expanding the result in powers of $c$ and truncating it to $O(c^{-2})$,  one can also show that 
\begin{align}\label{beta_phi}
\beta_\varphi=\frac{\ell}{\gamma_\text{tot}\,r}\Big(1-\frac{1}{c^2}\frac{G\,M}{r}\Big).
\end{align}
Moreover, the contravariant components of the four-velocity field are obtained as follows:
\begin{align}\label{u^mu}
\nonumber
&\Big(u^t,~u^r,~u^{\theta},u^{\varphi}\Big)=\bigg(c\,\gamma_\text{tot}\Big(1+\frac{1}{c^2}\frac{G\,M}{r}-\frac{1}{c^4}\big(\frac{G\,M}{r}\big)^2\Big),\\
&~V\Big(1+\frac{1}{2}\frac{V^2}{c^2}-\frac{1}{c^2}\frac{G\,M}{r}\Big),~0,~c\,\gamma_\text{tot}\,\omega+\frac{\ell}{r^2}\Big(1-\frac{1}{c^2}\frac{2\,G\,M}{r}\Big)\bigg).
\end{align}

\section{Coefficients of Wind equation}\label{app3}  

As discussed in the paper, we use continuity equation, radial and azimuthal momentum equations along with the EoS to calculate the gradient of flow variables, namely the wind equation ($dV/dr$) and the temperature gradient ($d C_{\rm s}/dr$). In the following, we describe the explicit forms of these equations and their coefficients. Here, we use a unit system as $G=M=c=1$.

\subsection{Semi-Relativistic limit (SR)} \label{SR_Wind}

\begin{align*}
    \frac{dV_{\rm SR}}{dr} = \frac{N_{\rm SR}}{D_{\rm SR}},
\end{align*}
where

\begin{align}
\nonumber
&N_{\rm SR} = N_{0_{\rm SR}} + \frac{C_{\rm s}^2}{2 (\Gamma -1) \Gamma  r^2}\Big[(\Gamma -1) \Big(5 r-2 (1-9 \Gamma )\Big) -5 C_{\rm s}^2 r- \frac{\ell}{2 r} \Big(\left(7 \Gamma ^2+\Gamma -8\right) \ell\\
& +24 (\Gamma +1) s\Big) + \frac{(\Gamma -1)}{r^3} \big(\ell^2-\ell r+r s\big) \big((\Gamma +4) \ell-6 s\big)\Big]  - \frac{d\Phi^{\rm SR}_{\rm eff}}{dr}, 
\end{align}
and
\begin{align}
& N_{0_{\rm SR}} = -\frac{C_{\rm s}^2 (\Gamma -1)}{4 \Gamma  \left(-C_{\rm s}^2+\Gamma +1\right)r^7} \Big(\Gamma  \ell^2+2 r (3 \Gamma +r)\Big)\\\nonumber
&\times\Big(\ell \Big[\ell
\big((\Gamma -4) \ell+(\Gamma -4) r^2-(\Gamma -4) r+6 s\big)+r s (\Gamma+12 r-10)\Big]-5 r^4+(3 \Gamma +2) r^3+6 r s^2\Big).
\end{align}

Similarly, the denominator takes the form,
\begin{align}
& D_{\rm SR} = -\frac{C_{\rm s}^2}{\Gamma V} + \frac{C_{\rm s}^4}{(\Gamma-1)V} - \frac{C_{\rm s}^2}{V}\bigg(\frac{3}{r} + \frac{\ell^2}{2r^2}\bigg) -\frac{(\Gamma -1) C_{\rm s}^2 \Big(\Gamma 
\ell^2+2 r (3 \Gamma +r)\Big)}{2 \Gamma  V \big(\Gamma
-C_{\rm s}^2 +1\big) r^2} + V.
\end{align}

Now, the coefficients for the temperature gradient equation are obtained as, 

\begin{align}
\nonumber
& C_{0_{\rm SR}} = \frac{C_{\rm s} (\Gamma -1)}{2  \left(-C_{\rm s}^2+\Gamma +1\right)r^5} \Big(\ell \Big[\ell \big((\Gamma -4) \ell+(\Gamma -4) r^2\\
&-(\Gamma -4)
r+6 s\big)+ s \,\,r (\Gamma +12 r-10)\Big]-5 r^4+(3 \Gamma +2) r^3+6 r s^2 \Big).
\end{align}

\begin{equation}
    C_{V_{\rm SR}} = -\frac{C_{\rm s} (\Gamma -1)}{V \left(-C_{\rm s}^2+\Gamma +1\right)}.
\end{equation}

\subsection{Semi-Newtonian limit (SN)} \label{SN_Wind}

In a similar way, we write the coefficients for the Semi-Newtonian case.

\begin{align}
\nonumber
&N_{0_{\rm SN}} = -\frac{C_{\rm s}^2 (\Gamma -1)}{\Gamma  (\Gamma +1) r^7}  \Big(\Gamma  \ell^2+2 r (3 \Gamma +r)\Big)\Big(\ell \big[\ell \big((\Gamma -4) \ell(\Gamma -4) r^2-(\Gamma -4) r+6 s\big)\\&+s\,\,r (\Gamma +12 r-10)\big]-5 r^4+(3 \Gamma +2) r^3+6 r s^2 \Big).
\end{align}

\begin{align}
\nonumber
&N_{\rm SN} = \frac{C_{\rm s}^2}{4\Gamma  r^5} \Big [\ell r^2 \big((7 \Gamma +8) \ell-24 s\big)+2 r^3 \big(5 r-2 (1-9 \Gamma )\big) + 2\big(\ell^2-\ell r+r s\big)\big((\Gamma +4) \ell-6 s\big)\Big] \\
&+\frac{1}{4}N_{0_{\rm SN}} - \frac{d\Phi_{\rm eff}^{\rm SN}}{dr}.
\end{align}

\begin{align}
D_{\rm SN} = V-\frac{C_{\rm s}^2 \big(\Gamma  \ell^2+2 r (3 \Gamma +r)\big)}{(\Gamma +1)  V r^2}.
\end{align}

\begin{align}
\nonumber
& C_{0_{\rm SN}} = \frac{C_{\rm s} (\Gamma -1)}{2 (\Gamma +1) r^5} \Big(\ell \big[\ell \big((\Gamma -4) \ell+(\Gamma -4) r^2-(\Gamma -4) r + 6 s\big)+ s \,\, r(\Gamma +12 r-10)\big] \\
&-5 r^4+(3 \Gamma +2) r^3+6 r s^2\Big).
\end{align}

\begin{align}
C_{V_{\rm SN}} = \frac{C_{\rm s}}{V} \frac{(1-\Gamma)}{(1+\Gamma)}.
\end{align}

\subsection{Non-relativistic limit (NR)} \label{NR_Wind}

\begin{align}
& N_{\rm NR}=\frac{d\Phi^{\rm N}_{\rm eff}}{dr} + \frac{6 + 10 r^2 + 3\Gamma(1-3\Gamma)+r(2+30\Gamma)}{2 (1+\Gamma)r^3},\\
& D_{\rm NR} = \frac{-2(r + 3\Gamma) C_{\rm s}^2}{(1+\Gamma)V r}+V.
\end{align}

\begin{align}
& C_{0_{\rm NR}} = \frac{(\Gamma-1)\,(2-5r+3\Gamma)C_{\rm s}}{2(1 + \Gamma)r^2 },\\
& C_{V_{\rm NR}}=\frac{C_{\rm s}(1-\Gamma)}{V(1+\Gamma)}.
\end{align}

\section{Discussion on angular momentum and disk luminosity} \label{discussion} 
\begin{figure}
    \centering   
    \includegraphics[width=275pt]{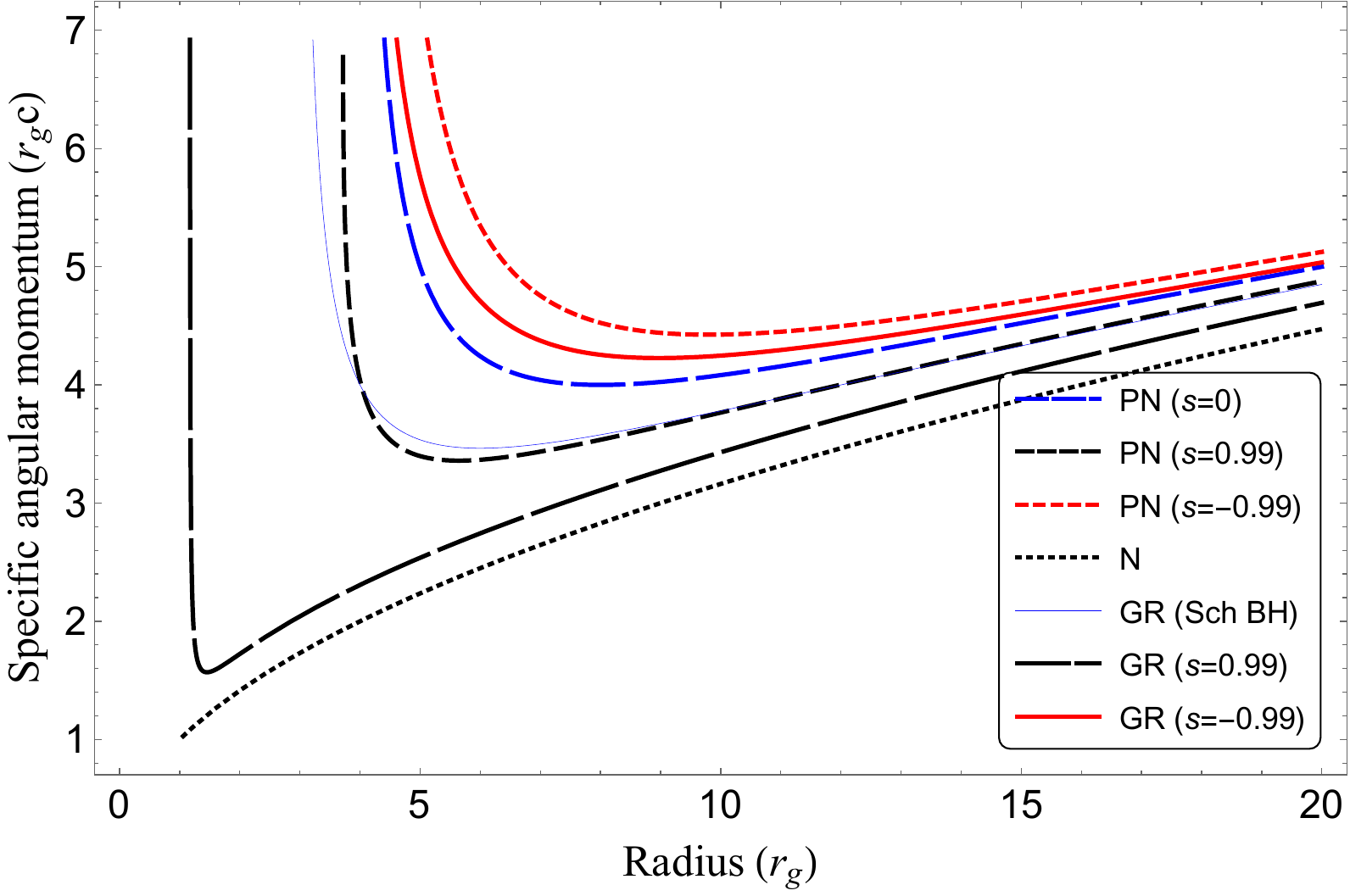}
    \caption{The relation between specific angular momentum $\ell$ and radius $r$ for the Keplerian circular motions for the Newtonian, post-Newtonian, and GR cases. Here, we illustrate three $s=-0.99$, $0$, and $0.99$ cases. This figure indicates that in the relativistic cases, \textit{i.e.,} the GR and post-Newtonian cases, $\ell_{\rm Kep}$ has a minimum. Therefore, unlike the Newtonian system, there can be a marginally stable circular orbit, which is a GR feature.}
    \label{fig2}
\end{figure}

We have investigated three possible systems in the previous sections. It is seen that for two of these cases, namely the SR and SN systems, the effective potential contains post-Newtonian corrections and deviates from the standard one. Then, it would be constructive to examine the Keplerian circular motions in these systems. To do so, we derive the angular momentum for which $d\Phi^{\rm PN}_{\rm eff}/dr=0$, \textit{i.e.,} $\ell^{\rm PN}_{\rm Kep}(r)$ where the subscript ``Kep'' denotes the Keplerian motion.  The results are summarized in Fig. \ref{fig2}. To compare our findings with the corresponding cases in GR, $\ell^{\rm GR}_{\rm Kep}(r)$ as well as $\ell^{\rm N}_{\rm Kep}(r)$ are inserted in this figure. For the exact GR angular momentum, we use the results given in Eqs. (2.99) and (2.101) of \cite{2008bhad.book.....K}\footnote{Note that $r_g$ in this reference is $\frac{2\,G\,M}{c^2}$.}.  

As seen, the post-Newtonian cases follow the GR results fairly well. As the spin of the central body increases, the minimum of $\ell_{\rm Kep}(r)$, corresponding to a marginally stable
circular orbit, moves toward the inner radii. This can be interpreted as a result of the frame-dragging \cite{2008bhad.book.....K}. Of course, this general relativistic behavior observed here goes beyond the validity regime of post-Newtonian mechanics, \textit{i.e.,} $r<r_{\text{PN}}$. However, it again demonstrates the effectiveness of this approximation.

\begin{figure}
    \centering
    \includegraphics[width=200pt]{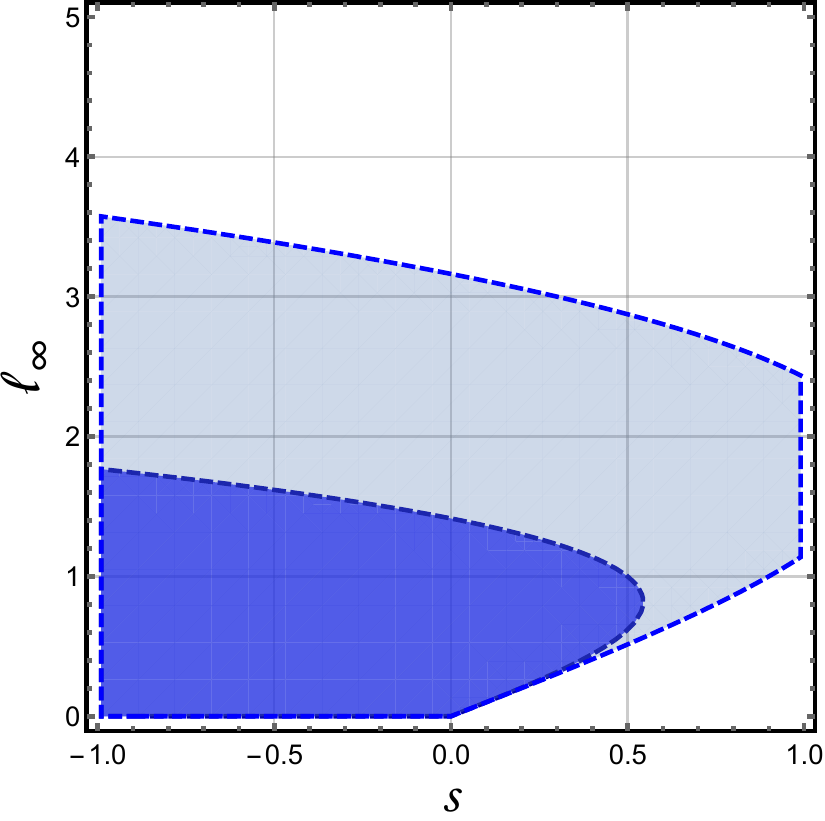}
    \caption{ The $s-\ell_{\infty}$ parameter space. The lighter (darker) blue area shows specific systems where angular momentum is continuously lost from $r_{\infty}$ to $r=10\,r_g$ ($r=2\,r_g$). Since only pure gravitational effects remove angular momentum up to these radii, these systems can be of interest. Note that in the present work, the effects of viscosity are ignored.}
    \label{fig3}
\end{figure}

Furthermore, in the SR and SN cases, we encounter an important gravitational relativistic effect that forces $\ell$ to vary with radius. So, let us return to Eq. \eqref{ell-SR}. Approximately solving this relation, we obtain that
\begin{align}
\ell_{\rm PN}=\ell_{\infty}+\frac{1}{r}\big(s-\ell_{\infty}\big)+\frac{\ell^3_{\infty}}{2\,r^2}.
\end{align}   
This post-Newtonian angular momentum is minimized at $r_{\rm min}=\frac{\ell_{\infty}^3}{\ell_{\infty}-s}$. We are interested in a system in which $r_{\rm min}$ takes place inside the region $r<r_{\text{PN}}$. In fact, in this case, we will have a special disk whose angular momentum decreases continuously up to the post-Newtonian radius.  Imposing the conditions $0<r_{\rm min}\leq 10$ provides us with the $s-\ell_{\infty}$ parameter space which represents this particular system. This area is depicted in light blue in Fig. \ref{fig3}. The bolder area corresponds to the stronger condition $0<r_{\rm min}\leq 2$ for which the system continuously loses angular momentum up to the Schwarzschild radius.

\begin{figure}
    \centering
    \includegraphics[width=275pt]{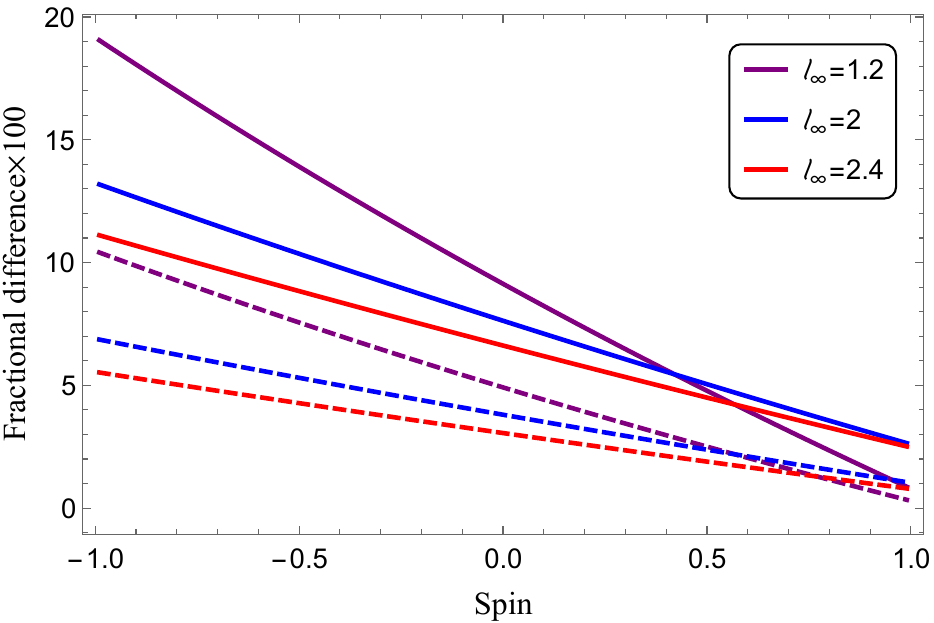}
    \caption{The fractional difference of angular momentum given in Eq. \eqref{frac-dif} in terms of spin for different values of $\ell_{\infty}$. The solid lines belong to the case $\big(\ell_{\rm PN,100}-\ell_{\rm PN,10}\big)/\ell_{\rm PN,10}$, while the dashed lines represent the case $\big(\ell_{\rm PN,20}-\ell_{\rm PN,10}\big)/\ell_{\rm PN,10}$. Here, the values of $\ell_{\infty}$ are selected so that the light blue area in Fig. \ref{fig3} covers them. }
    \label{fig4}
\end{figure}

In order to obtain how much angular momentum is removed from the system, we study the fractional difference
\begin{align}\label{frac-dif}
\bigtriangleup\ell_{\rm PN}=\frac{\ell_{\text{PN},\, i}-\ell_{\text{PN},10}}{\ell_{\text{PN},10}},
\end{align}
by choosing values of $\ell_{\infty}$ from Fig. \ref{fig3} that are suitable for any choice of $s$. Here, the numerical index shows the radius at which $\ell_{\text{PN}}$ is measured, \textit{i.e.,} $\ell_{\text{PN},10}=\ell_{\text{PN}}(r=10)$.
 We investigate two cases $i=100$ as well as $i=20$. Fig. \ref{fig4} reveals that in some cases, the pure gravitational relativistic effects remove a considerable amount of the angular momentum from the system. For instance, in the case with $\ell_{\infty}=1.2$ and $s=-1$, about $20\%$  of $\ell_{\text{PN},10}$ is lost from $r=100$ to $r=10$. Generally speaking, this release of angular momentum is more significant in retrograde disks with smaller $\ell_{\infty}$. Moreover, comparing the solid lines with the dashed ones, it can be seen that for almost all cases in this figure, about half of this value is taken away from the system by moving from $r=100$ to $r=20$, while the next half is removed by moving from $r=20$ to $r=10$. It is a reasonable result because relativistic effects become more effective in the inner parts of the disk, and consequently, they contribute more efficiently to angular momentum reduction.

However, the inflowing matter not only loses its angular momentum to fall onto the compact object, but has to release a significant amount of flow energy to reach near the compact body. In doing so, we consider the local potential energy released during the radiation motion from $r$ to $r-\Delta r$ in the vicinity of the central object following $\Phi^{\text{PN}}=-\frac{G\,M}{r}\Big[1+\frac{4}{3\,c^2}\Big(\frac{\ell^2}{r^2}-\frac{s\,\ell}{r^2}\Big)\Big]$, cf. Eq. \eqref{phi-PN}. We then have \cite{2008bhad.book.....K}
\begin{align}
\nonumber
\Delta L^{\text{PN}}_{\text{pot}}&=\Big(\Phi^{\text{PN}}(r)-\Phi^{\text{PN}}(r-\Delta r)\Big)\dot{M}\\
&\simeq \frac{G M\,\dot{M}}{r^2}\Big[1+\frac{4}{c^2}\Big(\frac{\ell^2}{r^2}-\frac{s\,\ell}{r^2}\Big)\Big]\Delta r.
\end{align}  
Crudely speaking, half of the potential energy $\Delta L^{\text{PN}}_{\text{pot}}$ will be radiated away. Therefore, the local radiation energy is 
\begin{align}\label{DeltaL}
\Delta L^{\text{PN}}_{\text{rad}}\simeq \frac{G M\,\dot{M}}{2\,r^2}\Big[1+\frac{4}{c^2}\Big(\frac{\ell^2}{r^2}-\frac{s\,\ell}{r^2}\Big)\Big]\Delta r,
\end{align}
and consequently, the disk luminosity is obtained as 
\begin{align}
\nonumber
L_{\text{d}}&=\int_{r_{\text{in}}}^{\infty}\frac{G M\,\dot{M}}{2\,r^2}\Big[1+\frac{4}{c^2}\Big(\frac{\ell^2}{r^2}-\frac{s\,\ell}{r^2}\Big)\Big]dr\\
&=\frac{G M\,\dot{M}}{2\,r_{\text{in}}}\Big[1+\frac{4}{3\,c^2\,r_{\text{in}}^2}\Big(\ell^2-s\,\ell\Big)\Big],
\end{align}
after integrating under a fixed $\ell$. 
Here, $r_{\text{in}}$ is the inner disk radius, which we choose as $r_{\text{in}}=r_{\text{PN}}$ in the current study.  
It is obvious that the post-Newtonian corrections that depend on $s$ and $\ell$ affect the luminosity of the disk. The negative (positive) term $\big(\ell^2-s\,\ell\big)$ results in a dimmer (brighter) disk.  

Regarding the relation $\Delta L^{\text{PN}}_{\text{rad}}=2\,\pi\,r\,\Delta r (2\,F)$, Eq. \eqref{DeltaL} also indicates that the flux of energy $F$ should be 
\begin{align}
F=\frac{G M\,\dot{M}}{8\,\pi\,r^3}\Big[1+\frac{4}{c^2\,r^2}\Big(\ell^2-s\,\ell\Big)\Big],
\end{align}
in the post-Newtonian framework. As seen, the standard portion of the energy flux changes as $1/r^3$ \cite{2008bhad.book.....K} while its relativistic correction is a function of $1/r^5$.

\section{Angular momentum transport in viscous accretion flows}\label{viscous}

The presence of viscosity in accretion disks is ubiquitous. However, in a convergent flow, the viscous time-scale ($t_{\rm vis}$) generally exceeds the infall time-scale ($t_{\rm in}$), particularly in the inner parts of the disk. As a consequence, the inflowing matter does not get sufficient time to transport the angular momentum outwards due to the differential motion, which results in an inviscid flow \cite[and references therein]{2019MNRAS.484.3209D}.

\begin{figure}[ht!]
    \centering
    \includegraphics[width=275pt]{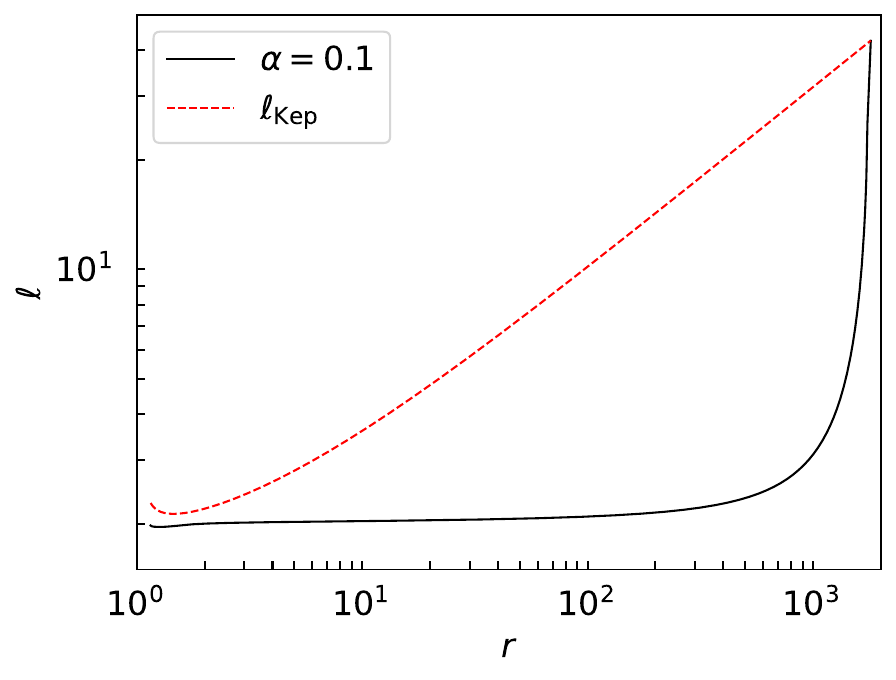} 
    \caption{Variation of flow angular momentum ($\ell$) of the general relativistic transonic accretion flow with radial coordinate ($r$) around a maximally rotating BH, with spin $s=0.99$ as shown by the solid curve. Here, the dashed curve represents the Keplerian angular momentum distribution ($\ell_{\rm Kep}$). See text for details given in Appx. \ref{discussion}.}
    \label{fig:l_r}
\end{figure}

In order to verify this behavior we take up a steady, axisymmetric and relativistic accretion disk around a maximally rotating Kerr BH (\textit{i.e.,} $s=0.99$), and include the effect of viscous stress tensor $\pi^{\mu\nu} = - 2 \nu \rho \sigma^{\mu\nu}$. Here, $\nu$ is the kinematic viscosity, $\rho$ is the mass density, and $\sigma^{\mu\nu}$ is the shear tensor following \cite{1973blho.conf..343N,2019MNRAS.484.3209D}. We use the similar approach to obtain a transonic accretion solution as described in Sec. \ref{sec5}. 

With this, in Fig. \ref{fig:l_r}, we compare the angular momentum distribution $(\ell)$ corresponding to a global transonic accretion solution. Here, the flow enters from the outer edge $r_{\rm edge}=2000$ of the accretion disk with an angular momentum $\ell = \ell_{\rm Kep}$ and energy $\mathcal{E}=1.001$. It is noteworthy that, even for a high viscosity parameter, $\alpha = 0.1$, the angular momentum variation remains quite insensitive up to $r \sim 1000$.
As we consider the low angular momentum flows, the circularization radius stays around few hundreds of $r_{\rm g}$. The above finding supports the assumption of inviscid nature of accretion flows in post-Newtonian formalism as well.

\end{document}